\documentclass[a4paper,11pt]{article}
\pdfoutput=1 

\usepackage{jheppub} 

\usepackage[T1]{fontenc} 
\usepackage{units}
\usepackage{subcaption}
\usepackage{graphicx}
\usepackage{float}
\usepackage[normalem]{ulem}
\RequirePackage[dvipsnames]{xcolor}

\newcommand{\Exp}[1]{\exp{\left\{#1\right\}}}

\title{\boldmath QCD antenna radiative spectrum in dense media within the Improved Opacity Expansion}

\renewcommand{\o}{\omega}
\newcommand{\re}{\text{Re}}

\newcommand{\nn}{\nonumber\\ }

\def\k{{\boldsymbol k}}
\def\q{{\boldsymbol q}}
\def\p{{\boldsymbol p}}
\def\n{{\boldsymbol n}}

\def\u{{\boldsymbol u}}
\def\x{{\boldsymbol x}}
\def\y{{\boldsymbol y}}
\def\z{{\boldsymbol z}}

\def\r{{\boldsymbol r}}

\def\Re{\text{Re} \, }

\newcommand{\cK}{\mathcal{K}}
\newcommand{\cP}{\mathcal{P}}

\newcommand{\cN}{\mathcal{N}}
\newcommand{\cR}{\mathcal{R}}
\newcommand{\cJ}{\mathcal{J}}
\newcommand{\cV}{\mathcal{V}}

\newcommand{\cI}{\mathcal{I}}

\def\bkappa{{\boldsymbol \kappa}}

\author[a]{Matvey V. Kuzmin}
\author[b,c,d]{Jo\~ao M. Silva}

\affiliation[a]{Physics Department, Lomonosov Moscow State University, 1-2 Leninskie Gory, Moscow 119991, Russia}

\affiliation[b]{Departamento de Física Teórica y del Cosmos, Universidad de Granada, Campus de Fuentenueva,
E-18071 Granada, Spain}

\affiliation[c]{Laboratório de Instrumentação e Física Experimental de Partículas (LIP), Av. Prof. Gama Pinto, 2, 1649-003 Lisbon, Portugal}

\affiliation[d]{Departamento de Física, Instituto Superior Técnico (IST), Universidade de Lisboa, Av. Rovisco Pais 1, 1049-001 Lisbon, Portugal}

\abstract{We compute the double differential inclusive spectrum for the emission of a soft gluon from a color-singlet 
$q\bar q$ pair traversing a dense QCD medium. Our results extend the existing literature by simultaneously incorporating both single hard and multiple soft gluon exchanges between the jet and the medium -- an essential ingredient for a complete phenomenological description of jet quenching. Using the Improved Opacity Expansion framework, we provide an analytically tractable treatment, reducing the full cross-section to a set of simple expressions. Our analysis demonstrates that rare hard (Molière) scatterings significantly modify the gluon spectrum at large angles, accelerating the loss of color coherence between the initial quarks. We further quantify whether the modifications are driven by an overall weakening of the interference term, or by more detailed modifications to the fragmentation pattern. Our results provide a direct input for phenomenological jet quenching studies, offering new insights into the role of color decoherence in QCD matter.
}

\begin{document} 
\maketitle
\flushbottom

\section{Introduction}~\label{sec:intro}
\indent The evolution of jets -- collimated particle cascades originating from an initial highly energetic parton -- in the QCD matter states produced in the aftermath of heavy-ion collisions is of much interest as a way to indirectly probe the underlying matter's properties~\cite{Apolinario:2022vzg, Mehtar-Tani:2013pia}. Since jets are produced and evolve in parallel to the bulk's expansion, they offer an unique opportunity to explore the real-time dynamics of the quark-gluon plasma (QGP) and construct a tomographic picture of it~\cite{Vitev:2004bh}. In order to achieve this, one has to understand how jet fragmentation is modified by the interactions with the background medium --- this class of effects is broadly referred to as jet quenching. In a perturbative setting and at high energies, it has been suggested that medium modifications can be computed by accounting for forward-scattering gluon exchanges between the jet partons and the medium \textit{quasi-particles} (see, e.g.,~\cite{Kovchegov:2012mbw}). Nonetheless, such computations are technically involved and one is mostly constrained to tree level processes with a small final state particle number. As a result, most of jet quenching phenomenology is solely driven by our understanding of how single-particle states' momenta broadens in matter and the associated radiative energy loss through \textit{bremsstrahlung}.

\begin{figure}[h!]
    \centering
	\includegraphics[height=3cm]{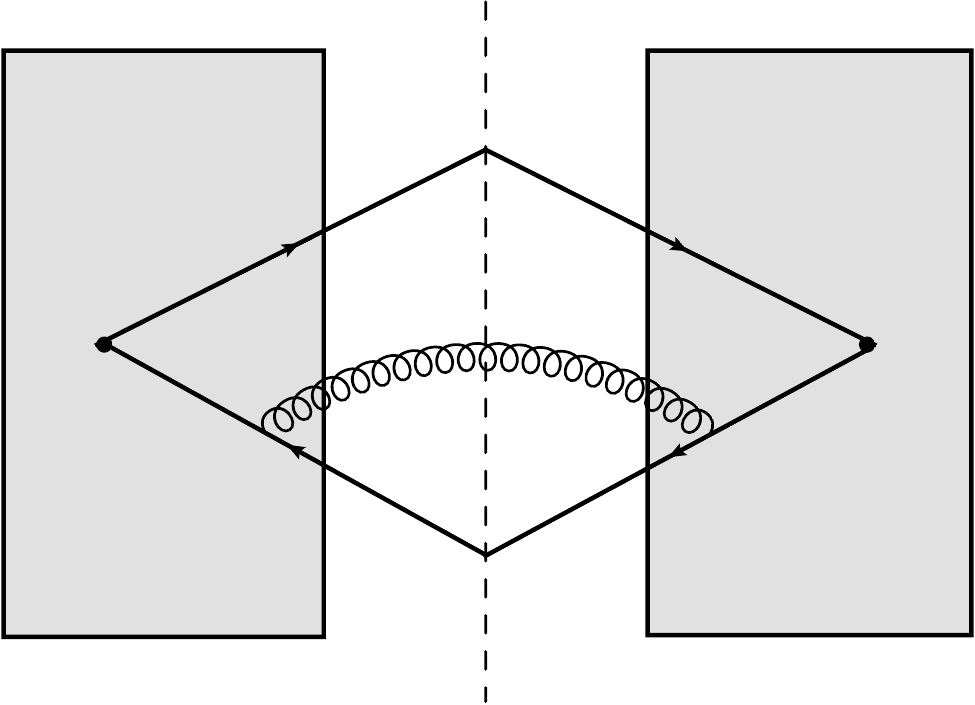}\hspace{1cm}\includegraphics[height=3cm]{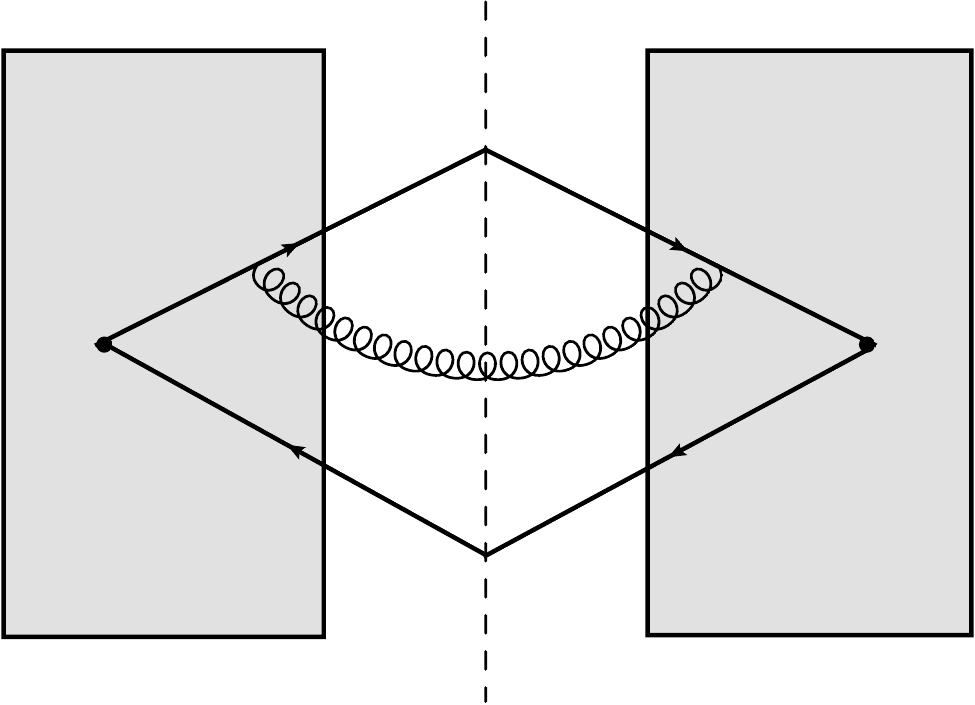}\hspace{1cm}\includegraphics[height=3cm]{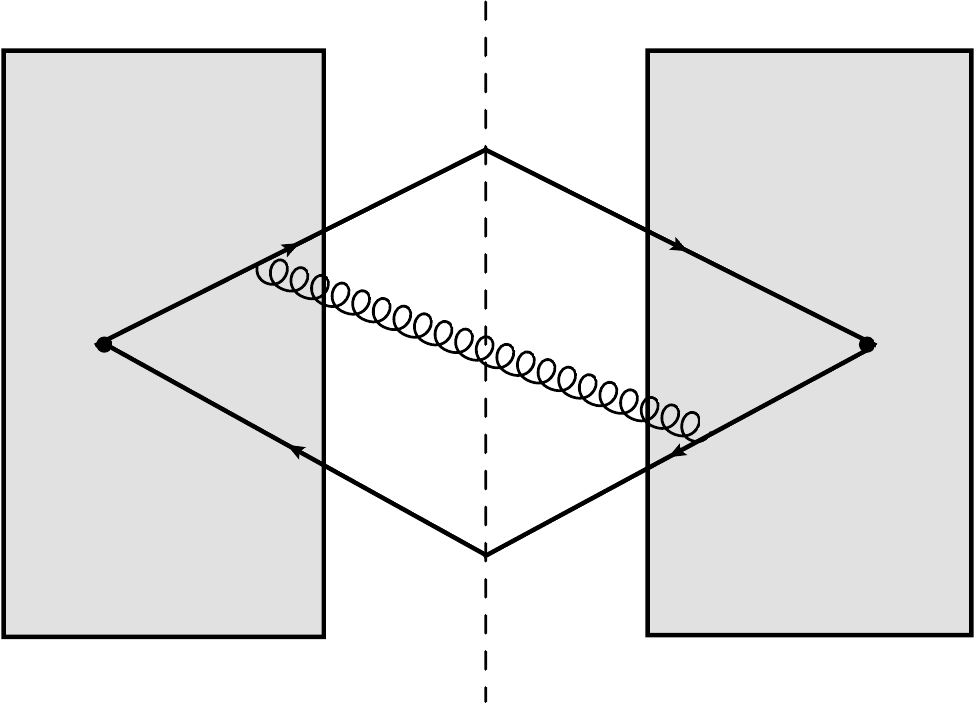}
	\vspace*{0mm}\caption{Three separate contributions to the QCD antenna gluon emission spectrum in Eq.~\eqref{eq:AntennaSpectrumDef}: two direct terms $\cR_q$ (\textbf{left}), $\cR_{\bar q}$ (\textbf{middle}) and interference term $\cJ$ (\textbf{right}). The gray box denotes the presence of a background medium with a finite extension.}\label{Fig:Antenna_Diagrams}
\end{figure}

In the past years significant effort has been put towards expanding our understanding of in-medium jet fragmentation beyond the above set-up, leading to studies on, e.g. finite energy branching processes~\cite{Apolinario:2014csa, Isaksen:2023nlr,Isaksen:2020npj}, multi-gluon production~\cite{Arnold:2024whj,Arnold:2022mby, Qian:2024gph}, finite formation time effects~\cite{Abreu:2024wka, Barata:2021byj, Apolinario:2024hsm}, detailed interactions with matter~\cite{Sadofyev:2021ohn}, among others. Along this line, substantial physical insight was gained by the computation of the single-gluon radiative spectrum off a \textit{classical} quark pair (antenna) source with a finite opening angle~\cite{Mehtar-Tani:2010ebp,Casalderrey-Solana:2012evi}, which in general can be written at leading order as
\begin{align}\label{eq:AntennaSpectrumDef}
	&(2\pi)^2\omega\frac{d\cN}{d\omega d^2\k} = \frac{\alpha_s C_F}{\omega^2}\left(\mathcal{R}_q+\mathcal{R}_{\bar q}-2\mathcal{J}\right)\,.	
\end{align}
Here $\mathcal{N}$ is the self-normalized single gluon inclusive cross-section, $\omega$ the frequency of the emitted gluon and $\k$ its transverse momentum. On the right hand side we have split the spectrum into three contributions: $\mathcal{R}_{q/\bar q}$ (the two leftmost diagrams in Fig.~\ref{Fig:Antenna_Diagrams}) denote the terms which, in particular gauges, can be interpreted as being the incoherent emission of the gluon starting from the quark/anti-quark; the interference pattern of the antenna is captured by $\mathcal{J}$ (rightmost diagram in Fig.~\ref{Fig:Antenna_Diagrams}). In the vacuum, and after angular averaging, as we show below, this term leads to a suppression of the spectrum for emissions angles $|\k|/\omega \approx \theta>\theta_{q\bar q}$, where $\theta_{q\bar q}$ is the opening angle of the antenna. This leads to the notion of angular ordering of QCD emissions in a cascade process~\cite{Marchesini:1987cf,Marchesini:1983bm}, at leading accuracy, as implemented in parton shower codes.

\begin{figure}[h!]
    \centering 
    \hspace{-0.5cm}\includegraphics[width=0.7\textwidth]{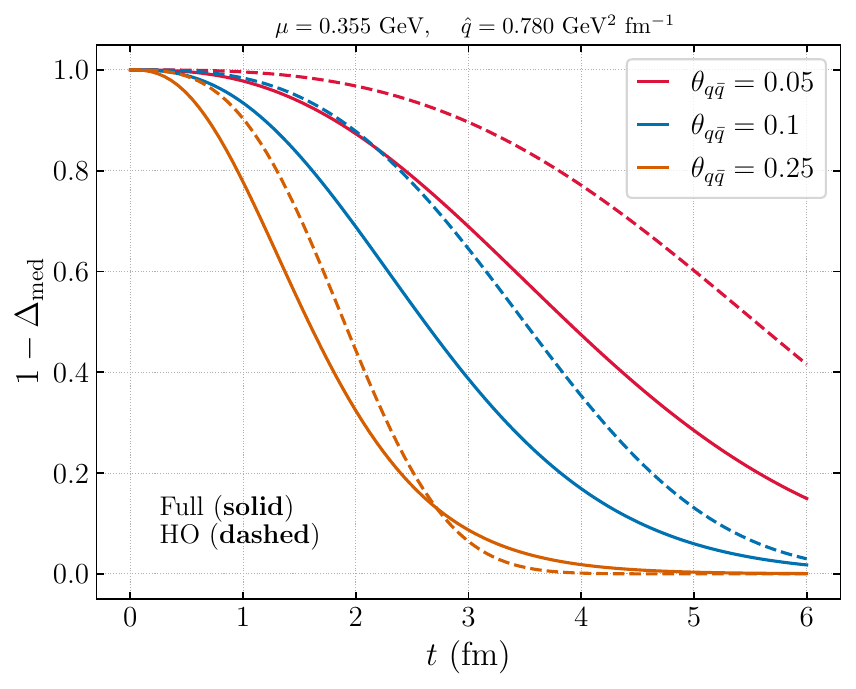}
    \vspace{-0.25cm}
    \\
     \caption{Time dependence of the decoherence factor $\left(1-\Delta_{\rm med}\right)$ defined in Eq.~\eqref{eq:CoherenceFDef} for three different opening angles:  $\theta_{q\bar q} = 0.05,\, 0.1\,,0.25$. The \textbf{solid} lines represent the decoherence factor calculated using the full dipole potential defined in Eq.~\eqref{eq:DipolePotentialGW}, while the \textbf{dashed} lines use the harmonic oscillator potential defined in the text just above Eq.~\eqref{eq:KLO}.
    }  \label{Fig:DecoherenceFactor} 
    \centering
\end{figure}

However, in the presence of a background medium, and in the limit of soft gluon radiation, one finds that the interference term satisfies $\mathcal{J}\propto (1-\Delta_{\rm med}(\theta_{q\bar q}))$~\cite{Mehtar-Tani:2012mfa}, which qualitatively modifies the form of the spectrum compared to the vacuum, where $\Delta_{\rm med}=0$. In Fig.~\ref{Fig:DecoherenceFactor} we show the evolution of this object with the propagation time, for several opening angles of the $q\bar q$ pair, for fixed values of the jet quenching parameter $\hat q$, Debye screening mass $\mu$ and using two approximations for the in-medium elastic scattering rate, which we discuss below. For all the curves, we observe that the longer the propagation in the medium, the smaller the interference term appears, leading to $\mathcal{J} \to 0$. This results in the loss of coherence of the radiative gluon spectrum~\cite{Mehtar-Tani:2012mfa,Mehtar-Tani:2011lic}, indicating that large antennas radiate as two independent color sources. Thus, gaining qualitative understanding as to how the interference term in Eq.~\eqref{eq:AntennaSpectrumDef} behaves as one accounts for more realistic descriptions of the jet-medium interactions might provide a better description of how the jet fragmentation pattern gets modified in heavy-ion collisions.

In this work we extend the computation of the in-medium gluon spectrum in Eq.~\eqref{eq:AntennaSpectrumDef} by accounting for soft and hard momentum exchanges with the medium, going beyond the harmonic/diffusion approximation previously used in~\cite{Mehtar-Tani:2012mfa}, incorporating the results for dilute matter considered in the companion paper~\cite{Mehtar-Tani:2011lic}. We gauge the size of these corrections to $\mathcal{J}$ in Fig.~\ref{Fig:DecoherenceFactor}, where the solid lines give the result including the full elastic scattering rate, while for the dashed ones we employ the harmonic approximation -- details of these calculation are discussed below -- accounting only for soft momenta exchanges with the medium. Thus, at the level of the decoherence factor, one can clearly observe that the inclusion of rare hard momentum exchanges leads to an accelerated incoherent gluon spectrum at small opening of the antenna, while at large angles the behavior becomes non-monotonic. Since these corrections are not negligible at the level of the decoherence factor $1-\Delta_{\rm med}$, this calls for a full analysis of $\mathcal{J}$ to gauge the effects of hard momentum exchanges in the antenna radiative spectrum.

The rest of this work is thus dedicated to incorporating the modifications shown in Fig.~\ref{Fig:DecoherenceFactor} to the full cross-section in Eq.~\eqref{eq:AntennaSpectrumDef}, i.e. in both $\mathcal{R}$ and $\mathcal{J}$. To begin, we start by reviewing the general form of the gluon spectrum emitted off a $q\bar q$ antenna in section~\ref{sec:antenna}, both in vacuum and in the medium. In section~\ref{sec:IOE}, we discuss how the Improved Opacity Expansion (IOE)~\cite{Mehtar-Tani:2019tvy,Barata:2020sav,Barata:2021wuf} framework can be used to incorporate both the single hard and multiple soft scattering regimes at the level of Eq.~\eqref{eq:AntennaSpectrumDef}. We note that in effect such an exercise has already been performed for the direct terms $\mathcal{R}_{q/\bar q}$ in~\cite{Barata:2021wuf}, while the interference term is first studied here. In section~\ref{sec:numerics}, we discuss the numeric results and their interpretation in light of the system's decoherence. We summarize our main findings in section~\ref{sec:conclusion}.

\section{QCD antenna: vacuum and in-medium spectrum}~\label{sec:antenna}
\indent The gluon radiation spectrum off a  QCD antenna in a color singlet state, i.e., produced from a virtual photon, with a fixed opening angle $\theta_{q\bar q}$ inside a longitudinally finite, homogeneous medium has been amply studied in the past for both dilute and dense matter~\cite{Mehtar-Tani:2010ebp,Mehtar-Tani:2011hma,Casalderrey-Solana:2011ule,Mehtar-Tani:2011lic,Mehtar-Tani:2011vlz,Armesto:2011ir,Mehtar-Tani:2012mfa,Casalderrey-Solana:2012evi,Calvo:2014cba}. In these works, the distribution of color charges in the medium follows a Gaussian form
\begin{align}\label{eq:two_point_correlator}
    \langle \mathcal{A}_a(t,\boldsymbol{x})\mathcal{A}_b(\bar t,\boldsymbol{y})\rangle = \delta_{ab}n(t)\delta(t-\bar t)\gamma(\y-\x)\, ,
\end{align}
where $\gamma(\x)$ is directly related to the in-medium elastic scattering rate and $n(t)$ is the number density of in-medium color sources, which in these works typically takes the form $n(t)=n_0\, \Theta(t < L)$, i.e., a static medium of length $L$. The initial photon is assumed to have a large virtuality, ensuring a vanishing formation time of the $q\bar q$ pair.\footnote{See~\cite{Barata:2021byj,Abreu:2024wka} for finite formation time effects.} This also implies that the produced quark and anti-quark are sufficiently energetic such that one can semi-classicaly approximate their trajectories, i.e., they are treated in the eikonal approximation (cf. e.g.~\cite{Altinoluk:2014oxa,Altinoluk:2015gia,Casalderrey-Solana:2007knd}). The gluon, on the other hand, exhibits transverse momentum broadening by interaction with the medium, undergoing transverse diffusion.\footnote{The longitudinal/transverse directions are defined with respect to the momentum of the initial photon.} The fact that all relevant dynamics is constrained to the transverse plane is a consequence of assuming the $+$ momentum component is dominant, translating into large light-cone energies. With these assumptions, the radiated gluon's spectrum differential in its energy $\omega$ and transverse momentum $\k$ is usually written as a sum of two direct components $\mathcal{R}_q$, $\mathcal{R}_{\bar q}$ and the interference term $\cJ$ and is presented in Eq.~(\ref{eq:AntennaSpectrumDef}).

In vacuum, this spectrum leads to the well-known characteristic pattern of antenna radiation~\cite{Dokshitzer:1991wu}
\begin{align}\label{eq:AntennaVacuum}    &(2\pi)^2\omega\frac{d\cN^{\text{vac}}}{d\omega d^2\k} = 4\alpha_s C_F\left(\frac{1}{\bkappa^2}+\frac{1}{\bar\bkappa^2}-2\frac{\bkappa\cdot\bar\bkappa}{\bar\bkappa^2\bkappa^2}\right)\,,
\end{align}
where $\bkappa = \k - x\p$ (with $x = \omega/E_{q}$) and $\bar{\bkappa} = \k - \bar{x}\bar{\p}$ (with $\bar x = \omega / E_{\bar q}$) correspond to the gluon's transverse momentum relative to the quark with momentum $\p$ and anti-quark with momentum $\bar \p$, respectively. The relative transverse momentum of the $q\bar q$ pair is given by $\delta\k = \bkappa-\bar{\bkappa}$ and its opening angle is directly related to this vector by $\delta \n = \delta\k/\omega \simeq \sin\theta_{q\bar q}$. A key feature of this spectrum is the phenomenon of \textit{angular ordering} of soft emissions. To see this, one starts by separating the total spectrum in a contribution containing the collinear divergence of the emission off the quark $d\cN_q \sim \cR_q - \cJ$ and another containing the one off the anti-quark  $d\cN_{\bar q} \sim \cR_{\bar q} - \cJ$.  By performing the azimuthal averaging and defining the polar angle of the gluon independently in each of the terms as relative to the quark or anti-quark directions, we get e.g. for the quark term
\begin{align}
    \omega\frac{d\cN_q^{\rm vac}}{d\omega d\theta}  = \frac{\alpha_s C_F}{\pi}\frac{\sin\theta}{1-\cos\theta}\Theta(\cos\theta-\cos\theta_{q\bar q})\,,
\end{align}
which implies the gluon is limited to be emitted inside the cone defined by the opening angle of the $q\bar q$ pair. For more details on vacuum spectrum and vacuum cascades see e.g. \cite{Mehtar-Tani:2011hma,Dokshitzer:1991wu,Dokshitzer:1982fh,Webber:1983if,Marchesini:1987cf,Dokshitzer:1987nm}.

To describe the gluon radiation off a color singlet $q\bar q$ antenna inside a medium, one performs a resummation of an arbitrary number of scatterings with the medium, such that for soft gluons (i.e. $\omega \ll E$, with $E$ the energy of the $q\bar q$ pair) the direct emission term for the quark is~\cite{Mehtar-Tani:2011hma,Casalderrey-Solana:2011ule,Mehtar-Tani:2012mfa, Mehtar-Tani:2011vlz}
\begin{align}\label{RqDef}
	\mathcal{R}_q = 2\Re\int_0^{\infty} dt_2 \,e^{-\varepsilon t_2}\int_0^{t_2} dt_1 \,e^{-\varepsilon t_1} \int_{\z}e^{-i\bkappa\cdot\z} \mathcal{P}(\infty,\z; t_2)\partial_{\y}\cdot\partial_{\z}\left.\cK(t_2,\z;t_1,\y)\right|_{\y = 0}\,,
\end{align}
and $\cR_{\bar q}$ is obtained from $\cR_q$ by replacing $\bkappa \rightarrow \bar\bkappa$. For the interference term one has
\begin{align}\label{eq:InterferenceDef}
	& \mathcal{J}= \Re \int_0^{\infty} dt_2 \,e^{-\varepsilon t_2}\int_0^{t_2} dt_1 \,e^{-\varepsilon t_1}\left(1-\Delta_{\rm med}(t_1)\right)e^{i\frac{\omega}{2}\delta\n^2 t_1}\nn
	& \hspace{0.5cm}\times\int_{\z} e^{-i\bkappa\cdot\z} \mathcal{P}(\infty,\z; t_2)\,(\partial_{\y}-i\omega\delta \n)\cdot\partial_{\z}\left.\cK(t_2,\z;t_1,\y)\right|_{\y = \delta\n t_1} + \rm{sym.}\,,
\end{align}
where the "+sym" contribution is found by replacing $\bkappa\rightarrow \bar\bkappa$ (which also implies $\delta \n \rightarrow-\delta \n$). We make explicit the adiabatic turn-off prescription~\cite{Wiedemann:2000za} by including $e^{-\varepsilon t}$ for each time integration, preventing vacuum-like radiation at arbitrarily large times. The limit $\varepsilon\rightarrow 0$ is implicit in Eqs.~\eqref{RqDef} and~\eqref{eq:InterferenceDef} and throughout the remaining of this work. Finally, note that in the limit of $\delta\n\rightarrow 0$, the interference term $\cJ$ exactly cancels out both direct terms. This property is understood as follows: if the $q \bar q$ pair has negligible spatial separation, corresponding to a vanishing opening angle, the pair behaves as a single color-neutral projectile, and thus cannot radiate.
 
The functions $\mathcal{P}$ and $\cK$ are typically referred to as the broadening and emission kernels, respectively. The former contribution is given explicitly by 
\begin{align}\label{eq:BKernelDef1}
    & \mathcal{P}(\infty,\z;t_2) = \Exp{-\int_{t_2}^{\infty} ds\, \cV(\z,s)}\,,
\end{align}
while the latter one is the solution to the following $2+1$D Schrödinger equation
\begin{align}\label{eq:EkernelDef}
    &\left[i\frac{\partial}{\partial t} + \frac{\partial^2_{\x}}{2\o}+i \cV(\x, t)\right]\cK(\x,t;\y,t_1) = i \delta^2(\x-\y)\delta(t-t_1)\, ,
\end{align}
which can be equivalently represented by a 2D non-relativistic quantum mechanical path integral
\begin{align}\label{eq:general_radiative_kernel}
    & \cK(t_2,\z;t_1,\y) = \int_\y^\z \mathcal{D}\r\Exp{\int_{t_1}^{t_2} ds \, \left(i\frac{\omega}{2}\dot\r^2 - \cV(\r,s)\right)}\,.
\end{align}
The function $\mathcal{V}(\x,s)$ can be understood as an in-medium scattering potential, which controls the interaction between the medium and each parton. It can be written in terms of a model-dependent single-source scattering potential
\begin{align}\label{DipolePotentialDef}
    \cV(\x,t) =  C_Ag^2 n(t)(\gamma(0) - \gamma(\x)) = C_A n(t)\int_{\q} |v(\q^2)|^2
    \left(1-e^{i\q\cdot\x}\right)\,,
\end{align}
where $C_A$ is the quadratic Casimir in adjoint representation.\footnote{Note that multiple notations for the dipole potential $\cV(\x)$ and for $\gamma(\x)$ exist in the literature. The important point is that one is consistent between the definition of the two-point correlator in Eq.~\eqref{eq:two_point_correlator}, the dipole potential definition in, for instance, Eq.~\eqref{eq:DipolePotentialGW} and the kernel definition in Eq.~\eqref{eq:general_radiative_kernel}.} While there has been significant progress in extending the definition of this potential to include, for instance, transverse matter gradients, anisotropies~\cite{Sadofyev:2021ohn, Hauksson:2021okc,Barata:2022utc,Barata:2022krd,Fu:2022idl,Barata:2023qds,Kuzmin:2023hko,Hauksson:2023tze,Ke:2024emw,Barata:2024bqp,Barata:2024xwy,Barata:2025htx} and flow~\cite{Sadofyev:2021ohn,Antiporda:2021hpk,Andres:2022ndd,Kuzmin:2023hko, Bahder:2024jpa,Kuzmin:2024smy} effects, in the current setup we focus on a static, homogeneous and isotropic medium. Thus, in what follows, we employ the Gyulassy-Wang (GW) in-medium model \cite{Wang:1992qdg, Gyulassy:2000er}, noting that the main results can be readily extended to a broader class of models (see, e.g., \cite{Barata:2020rdn, Mehtar-Tani:2019tvy, Barata:2021wuf} for a discussion of universality). Within this model, the medium is described as a set of stochastic, static color sources interacting via Yukawa-like potentials
\begin{align}
    v^{\text{GW}}(\q^2) = -\frac{g^2}{\q^2+\mu^2}\,,
\end{align}
where $\mu = g T$, with $T$ the medium temperature, is the GW screening mass. This leads to a scattering potential of the form
\begin{align}\label{eq:DipolePotentialGW}
      \cV^{\text{GW}}(\x) = \frac{\hat{q}_0}{\mu^2}\left [ 1-|\mu \x|\,K_{1}\left(|\mu \x|\right)\right ]  \, ,
\end{align}
where $\hat q_0 = 4\pi\alpha_s^2 C_R n(t)$ is the bare jet quenching parameter, quantifying the mean transverse momentum accumulated per mean free path by a colored projectile traversing the medium. 

Finally, one other important component of the interference spectrum in Eq.~\eqref{eq:InterferenceDef} is the survival probability, often referred to as the medium decoherence factor \cite{Mehtar-Tani:2010ebp,Mehtar-Tani:2011vlz, Mehtar-Tani:2011hma,Mehtar-Tani:2011lic}
\begin{align}\label{eq:CoherenceFDef}
	& 1-\Delta_{\rm med}(t_1) = \Exp{-\int_0^{t_1}ds\, \cV(\delta\n s)}\, .
\end{align}
This factor quantifies the rate of color decoherence of the $q\bar q$ pair prior to gluon emission as a result of color exchanges with the medium. In the limit of vanishing gluon energy $(\omega \rightarrow 0)$~\cite{Mehtar-Tani:2010ebp,Mehtar-Tani:2011hma} this factor is multiplicative and it admits a simple interpretation --- while vacuum radiation is angular ordered, purely medium-induced radiation is \textit{anti-angular ordered}. In the case where an arbitrary number of scatterings is taken into account~\cite{Mehtar-Tani:2011hma}, $\Delta_{\rm med}$ varies between $0$ (a completely transparent medium), where it corresponds to the angular-ordered gluon spectrum in vacuum, and $1$ (a completely opaque medium), where it implies a totally \textit{incoherent} gluon spectrum, i.e., independent radiation off the quark and anti-quark. The latter case implies a ‘loss of memory’ of color connections between quark and anti-quark after interacting with the medium~\cite{Mehtar-Tani:2012mfa}.

When considering a gluon with finite energy~\cite{Casalderrey-Solana:2011ule, Mehtar-Tani:2011vlz, Mehtar-Tani:2011lic, Mehtar-Tani:2012mfa}, there is a more complex structure of the transition between two regimes dominated by either coherence or incoherence of the $q\bar q$ antenna. This transition is parametrically governed by an interplay between the maximum dipole size $r_{\perp} = \theta_{q\bar q}L$ and a medium-induced scale. In the case of a finite, dense medium~\cite{Mehtar-Tani:2012mfa}, and in the harmonic or multiple soft scattering approximation, that we will define in Section~\ref{subsec:BroadeningIOE}, the decoherence factor takes the form
\begin{align}\label{eq:harmonic_deltamed}
    \Delta_{\rm med}(t) = 1 - \Exp{-\frac{1}{12}Q_{s0}^2 r_{\perp}^2\left(\frac{t}{L}\right)^3}
\end{align}
such that the medium-induced scale $Q_{s0}^2 = \hat q_0 L$ can be identified as a (bare) saturation scale. It follows that two regimes~\cite{Mehtar-Tani:2012mfa} can be identified: \textit{dipole} where $r_{\perp} < Q_{s0}^{-1}$, and  \textit{partial decoherence} when $r_{\perp} > Q_{s0}^{-1}$. In the former regime, the decoherence factor is small and reads
\begin{align}
    \Delta_{\rm med}(L) \sim Q_{s0}^2 r_{\perp}^2\,,
\end{align}
resulting only in partial decoherence of the spectrum and, in the case of vanishing dipole size $r_{\perp}$ or saturation scale $Q_{s0}$, total color transparency. This interplay can also be formulated in terms of time estimates, where the dipole condition can be translated into $t_{d0} \sim (\hat q_0 \theta_{q\bar q}^2)^{-1/3} > L$, i.e., the average time it takes for the $q\bar q$ to be decohered by the medium is typically larger than the medium's length. In the opposite regime, the quark and anti-quark, on average, decohere due to interactions with the medium after a short time ($t_{d0} < L$), causing $\Delta_{\rm med}(L)$ to saturate at unity, in the limit of sufficiently small $t_{d0}$. In this case, the interference term is suppressed and the quark and anti-quark radiate, predominantly, as independent color sources.

It should be emphasized that the harmonic approximation employed to obtain the simple expression for $\Delta_{\rm med}$ in Eq.~\eqref{eq:harmonic_deltamed} is typically only valid for sufficiently soft ($\omega \ll \omega_{c0} = \frac{\hat q_0 L^2}{2}$) and collinear ($\k \ll Q_{s0}$) gluon emissions. Hence, for more energetic gluons 
the radiation spectrum becomes sensitive to single hard scatterings with the medium, more well described within the opacity expansion approach~\cite{Mehtar-Tani:2011lic}. A discussion on how to incorporate the regimes of multiple soft scattering and of a single hard scattering~\footnote{ See~\cite{Isaksen:2022pkj} where the two regimes and the Bethe-Heitler regime are included in the radiative spectrum.} to describe in-medium color coherence is presented in what follows.

\section{Review of the Improved Opacity Expansion}~\label{sec:IOE}
\indent Although the dipole potential in Eq.~\eqref{eq:DipolePotentialGW} has a known closed-form analytical expression, its direct implementation into subsequent analytical calculations of the emission spectrum contributions in Eqs.~\eqref{eq:BKernelDef1} and~\eqref{eq:EkernelDef} proves challenging. Consequently, simplifying assumptions regarding the interaction are typically employed to allow for an analytical treatment of the spectrum, including the harmonic or multiple soft scattering approximation alluded to in Section~\ref{sec:antenna}. In~\cite{Mehtar-Tani:2019tvy,Mehtar-Tani:2019ygg,Barata:2020rdn,Barata:2020sav,Barata:2021wuf},\footnote{See also e.g.~\cite{Adhya:2024nwx, Mehtar-Tani:2024jtd} for some more recent developments and applications of this strategy.} the authors developed a strategy to systematically calculate corrections to this approximation, such that the effect of few harder medium-parton scatterings can be included in a analytically driven treatment.
This framework was dubbed the Improved Opacity Expansion (IOE), and builds on the seminal work by Molière~\cite{Moliere+1948+78+97}. In the following section, we will briefly review the IOE approach using two established examples (cf. \cite{Barata:2021wuf} for a more detailed review), before applying it to the calculation of the radiation spectrum of a color singlet $q\bar q$ antenna, which constitutes the primary objective of this work.

\subsection{Transverse momentum broadening within the IOE}\label{subsec:BroadeningIOE}
\vspace*{-3mm}~\\
\indent We begin by reviewing the simplest effect arising from jet-medium interactions, momentum broadening, whereby a single highly energetic particle propagates through a dense medium, undergoing multiple elastic scatterings. The probability for this particle to acquire a given transverse momentum $\k$ during a time $L$ is described by the broadening kernel $\cP(L,\k)$. While this kernel has a particularly elegant form when Fourier transformed,
\begin{align}\label{eq:BKernalDef}
    \cP(L,\z) = e^{- \cV(\z)L},
\end{align}
a closed-form analytical solution for $\cP(L,\k)$ for the GW potential in Eq.~\eqref{eq:DipolePotentialGW} does not exist. Taking the limit of small dipole size, we can expand the effective potential in powers of $\mu^2\z^2 \ll 1$,  resulting in a dominant term of the form
\begin{align}\label{eq:small_dipole_size}
    \cV(\z) \simeq \frac{1}{4}\hat{q}_0 \z^2 \log\frac{1}{\mu^2_{*}\z^2}\,,
\end{align}
where $\mu^2_{*} = \frac{1}{4} e^{-1+2\gamma_E}\mu^2$ is the universal mass, directly related to the Debye screening mass $\mu$. However, obtaining an analytical solution for $\cP(L,\k)$ remains challenging, thus the harmonic oscillator (HO) approximation is typically employed, completely ignoring the logarithmic dependence~\cite{Baier:1998yf, Zakharov:1998wq}. In this case the broadening kernel reads:
\begin{align}\label{eq:HOP}
    \cP^{\text{HO}}(L,\k) = \int_{\x} e^{-i\k\cdot\x}e^{- \frac{1}{4}Q_{s0}^2 \z^2} = \frac{4\pi}{Q_{s0}^2}e^{-\frac{\k^2}{Q_{s0}^2}}\,,
\end{align}
with $Q_{s0}$ the bare saturation scale introduced above. 
Despite its widespread phenomenological applications, this approximation is valid only in the multiple soft scattering (MS) regime, i.e., when the opacity parameter is relatively large ($\chi\sim Q_{s0}^2/\mu^2_{*} \gg 1$). It fails to accurately describe the dilute medium case ($\chi\ll 1$), where the single hard scattering (SH) regime dominates the distribution (see \cite{Barata:2020rdn} for further details). The goal of the IOE is to capture the key features of both regimes. First, we rewrite the dipole potential as
  
\begin{align}\label{eq:IOEPotential}
    \cV(\z) \simeq \frac{1}{4}\hat{q}_0 \z^2 \log\frac{1}{\mu^2_{*}\z^2} = \frac{1}{4}\hat{q}_0 \z^2 \left(\log\frac{Q_b^2}{\mu^2_{*}}+\log\frac{1}{Q_b^2\z^2}\right) \, .
\end{align}
Then the IOE prescription implies choosing a matching scale $Q_b$ such that the second term in the expansion in Eq.~\eqref{eq:IOEPotential} can be treated as a small correction to the leading-order (LO) harmonic oscillator term, allowing the potential to be expressed as
\begin{align}\label{eq:dipole_potential_separation}
    \cV(\z) = \cV^{\text{LO}}(\z)+\delta\cV(\z)\,, \quad \delta\cV \ll \cV \, .
\end{align}
To satisfy this condition, we require the momentum scale hierarchy $Q_b^2 \gg \mu_{*}^2$. This form of the dipole potential enables the expansion of the final distribution as a perturbative series in coordinate space
%
\begin{align}\label{eq:BKernelExpansion}
    \cP(L,\z) = \cP^{\text{LO}}(L,\z)+\cP^{\text{NLO}}(L,\z)+... 
\end{align}
where
\begin{align}\label{eq:LO_NLO_}
    \cP^{\rm LO}(L,\z) &= e^{-\frac{1}{4}Q_s^2\z^2}\, ,
    \nn 
     \cP^{\rm NLO}(\z,L)  &= \frac{1}{4}\hat q_0 \z^2 \log \frac{1}{Q_b^2\z^2}e^{-\frac{1}{4}Q_s^2\z^2}\, ,
\end{align}
and in momentum space
\begin{align}\label{eq:BKernel_momspace}
    \notag&\cP(L,\k)  =\int_{\x} e^{-i\k\cdot\x}e^{-\frac{1}{4}\x^2Q_s^2}e^{-\frac{1}{4}\x^2 Q_{s0}^2\log(\frac{1}{\x^2 Q_b^2})}
    \nn&\hspace{1cm}=\int_{\x} e^{-i\k\cdot\x}e^{-\frac{1}{4}\x^2Q_s^2}\sum_{n}\frac{(-1)^nQ_{s0}^n}{4^n n!}\x^{2n}\log^n\left(\frac{1}{\x^2 Q_b^2}\right) 
    \nn
    &\hspace{2cm}= \cP^{\text{LO}}(L,\k)+\cP^{\text{NLO}}(L,\k)+...
\end{align}
where we have defined the \textit{effective} saturation scale $Q^2_s=\hat{q} L$ with an \textit{effective} jet quenching parameter $\hat{q}=\hat q_0\log\frac{Q^2_b}{\mu_{*}^2}$. The integrals in Eq.~\eqref{eq:BKernel_momspace} above can be evaluated analytically, such that the broadening kernel up to NLO is given by~\cite{Barata:2020rdn}
\begin{align}\label{eq:BroadeningExpanded}
    \cP^{\text{LO}}(L,\k)+\cP^{\text{NLO}}(L,\k) = \frac{4\pi}{Q_{s}^2}e^{-x} -\frac{4\pi}{Q_{s}^2}\lambda\left(1-2e^{-x}+(1-x)\left[\text{Ei}(4x)-\log 4x\right]\right)\,,
\end{align}
with $x = \k^2/Q_s^2$ and $\lambda = \hat{q}_0/\hat{q}=1/\log\frac{Q^2_b}{\mu_{*}^2}\ll 1$ being an expansion parameter.\footnote{The exponential integral is defined as $\text{Ei}(x) = \int_{-\infty}^{x}dt\, \frac{e^t}{t}$.} 
While there is a priori no 
unique, model-independent condition that determines the matching scale $Q_b$, since it is part of the definition of $\hat q$,  one can define it through
the transcendental equation 
\begin{align}\label{eq:QbCondition}
    Q_b^2(L) = Q_s^2(L) = \hat{q}_0 L\log\frac{Q^2_b}{\mu_{*}^2}\,.
\end{align}
 This prescription has been previously used by Molière in~\cite{Moliere+1948+78+97} (see also~\cite{Barata:2020rdn, Barata:2021wuf}) and it corresponds to the average transverse momentum acquired by a parton traveling the whole extent of the medium. By further expanding the LO+NLO kernel in Eq.~\eqref{eq:BroadeningExpanded} for either large ($\k^2 \gg Q_s^2$) or small ($\k^2 \ll Q_s^2$) momentum transfer, one obtains the correct broadening solutions for the single hard scattering and multiple soft scattering regimes, respectively~\cite{Barata:2020rdn}. 

\subsection{Radiation spectrum within the IOE}\label{subsec:RadiationIOE}
\vspace*{-3mm}~\\
\begin{figure}
    \centering
	\includegraphics[height=4cm]{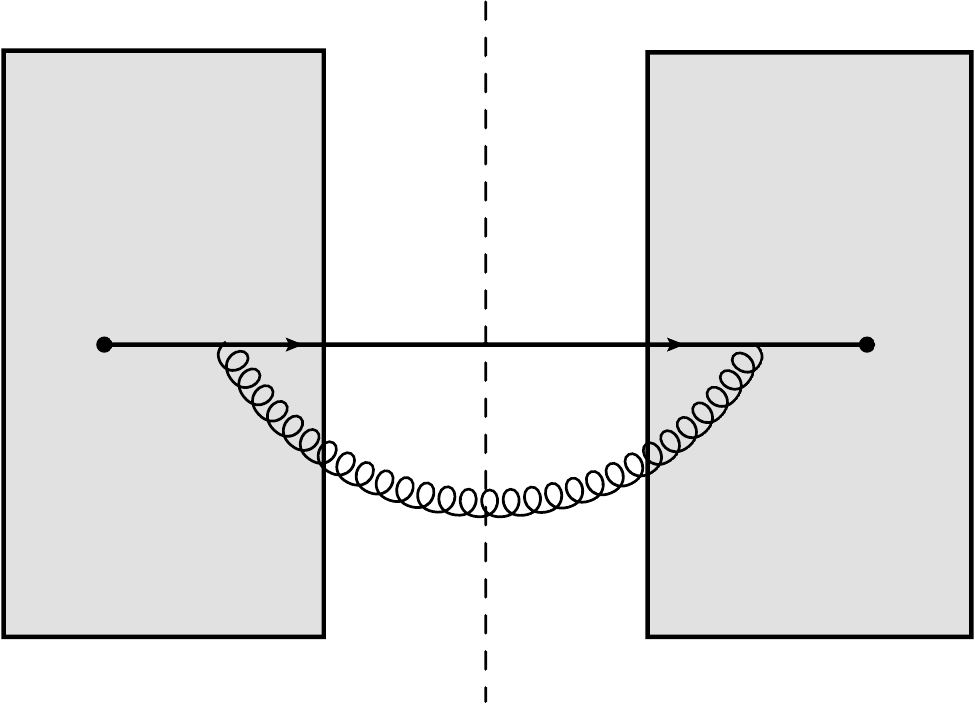}
	\vspace*{0mm}\caption{Diagram corresponding to gluon emission off a single quark inside a medium, illustrating the calculation in Eq.~\eqref{eq:QuarkGluonSpectrum}.}\label{Fig:QuarkEmission}
\end{figure}
\indent After calculating transverse momentum broadening within the IOE, the next natural step is to examine the in-medium radiation spectrum of a soft gluon with energy $\o$ and transverse momentum $\k$, emitted by a single highly energetic quark with energy $E$ (see Fig.~\ref{Fig:QuarkEmission}). This spectrum can be compactly written as (see e.g.~\cite{Baier:1996sk, Baier:1996kr,Zakharov:1997uu,Arnold:2002ja,Wiedemann:2000za,Blaizot:2015lma,Barata:2021wuf})
\begin{align}\label{eq:QuarkGluonSpectrum}
    \notag&
    (2\pi)^2\o\frac{d \cI}{d\o d^2\k} = \frac{\alpha_s C_F}{\o^2} \re \int_0^\infty dt_2 \,e^{-\varepsilon t_2}\int_0^{t_2}dt_1 \,e^{-\varepsilon t_1}
    \\
    &\hspace{1cm}\times\int_{\z}e^{-i\k\cdot\z}\cP(\infty,\z;t_2)\partial_\y \cdot\partial_\z \cK(t_2,\z;t_1,\y)|_{\y=0}\,,
\end{align}
where we recall the definitions in Eqs.~\eqref{eq:BKernelDef1} and~\eqref{eq:EkernelDef} of the broadening and emission kernels, respectively. Evidently, the distribution above is directly related to the direct contribution $\cR_q$ to the antenna spectrum defined in Eq.~\eqref{RqDef}, serving as an important building block to understand the full result. This spectrum includes both \textit{medium-induced} and \textit{vacuum} contributions. The pure vacuum radiation off the hard emitter is described by
\begin{align}\label{eq:QuarkGluonSpectrumVacuum}
    (2\pi)^2\o\frac{d \cI^{\rm vac}}{d\o d^2\k}= \frac{4\alpha_s C_F}{\k^2}\,, 
\end{align}
to describe pure medium-induced gluon radiation, one subtracts this contribution from the total spectrum in Eq.~\eqref{eq:QuarkGluonSpectrum}.

As it was mentioned in Section~\ref{sec:antenna}, the emission kernel $\cK(t_2,\x;t_1,\y)$ is the solution of a $2+1$D Schrödinger equation (see Eq.~\eqref{eq:EkernelDef}) with the dipole potential $\cV(\x, t)$ as a source for jet-medium interactions. An analytical closed solution of this equation for the GW model in Eq.~\eqref{eq:DipolePotentialGW} cannot be found. Thus, it is instructive to consider several simplifications that allow for analytical treatment of the spectrum.
\newcounter{interactions}[section]
\begin{itemize}
\item The simplest scenario is the one with \textbf{no interactions}, i.e., $\cV(\x, t)=0$, resulting in the vacuum kernel
\begin{align}\label{eq:KVac}
   &\cK_0(t_2,\x;t_1,\y) = \frac{\o}{2\pi i (t_2-t_1)}\exp\left \lbrace i\frac{\o(\x-\y)^2}{2(t_2-t_1)}\right\rbrace\,.
\end{align}
This kernel describes the gluon emission outside of the medium. By plugging this kernel in Eq.~\eqref{eq:QuarkGluonSpectrum} one naturally obtains the vacuum spectrum in Eq.~\eqref{eq:QuarkGluonSpectrumVacuum}. 

\item Following the small dipole size expansion in Eq.~\eqref{eq:small_dipole_size}, another 
relevant example is the \textbf{harmonic oscillator (HO) approximation} which, as mentioned in the previous section, implies neglecting the remaining logarithmic dependence, implying a dipole potential that is quadratic in distance, i.e., $\cV^{\text{HO}}(\z) = \hat{q}_0\,\z^2/4$, where $\hat q_0$ is the \textit{bare} jet-quenching parameter. Within this setup, the Schrödinger Eq.~\eqref{eq:EkernelDef} can be analytically solved:
%
\begin{align}\label{eq:KLO}
    	& \cK^{\text{HO}}(t_2,\z;t_1,\y) =\frac{\omega}{2\pi i S_{21}}\Exp{\frac{i\omega}{2S_{21}}[C_{21}(\z^2+\y^2)-2\z\cdot\y]}\,,
\end{align}
with shorthand notations 
\begin{align}
	& S_{21} = \frac{\sin (t_2-t_1)\Omega}{\Omega}\,,\quad C_{21} = \cos (t_2-t_1)\Omega\,, \quad \Omega = \frac{1-i}{2}\sqrt{\frac{\hat q_0}{\omega}}\,. 
\end{align}
Although in this paper we focus on the static finite medium case, these functions can be adapted for evolving medium profiles (see e.g. \cite{Salgado:2003gb,Baier:1998yf,Arnold:2008iy,Adhya:2019qse}).
Taking the form of the emission kernel in Eq.~\eqref{eq:KLO} and the coordinate space version of the broadening kernel in Eq.~\eqref{eq:HOP}, one can evaluate the $\omega$ and $\k$ differential gluon spectrum in the harmonic oscillator approximation. As discussed in the case of broadening of a single parton in Section~\ref{subsec:BroadeningIOE}, a downfall of the HO limit is that it fails to describe the regime where a single hard momentum exchange with the medium is relevant. This is the case when describing an emitted gluon with large transverse momentum $\k \gg Q_{s0}$ and relatively large energy $\omega \gg \omega_{c0}$, where $\omega_{c0} = \frac{\hat q_0 L^2}{2}$.

\item Finally, in the \textbf{Improved Opacity Expansion} one obtains a perturbed HO result. Exactly like for the process of transverse momentum broadening (see Eqs.~\eqref{eq:IOEPotential} and~\eqref{eq:dipole_potential_separation}), we start by splitting the potential in the small dipole size approximation into two contributions
\begin{align}
    \cV(\z)  = \frac{1}{4}\hat{q}_0 \z^2 \left(\log\frac{Q_r^2}{\mu^2_{*}}+\log\frac{1}{Q_r^2\z^2}\right)= \cV^{LO}(\z)+\delta\cV(\z)\,, \quad \delta\cV \ll \cV\,,
\end{align}
where $\cV^{\rm LO}$ gives rise to the HO solution in Eq.~\eqref{eq:KLO}, but now with the \textit{bare} jet quenching parameter $\hat q_0$ replaced by the \textit{effective} jet quenching parameter $\hat{q} = \hat{q}_0 \log\frac{Q_r^2}{\mu_{*}^2}$. The matching scale for the emission kernel does not necessarily coincide with $Q_b$, defined for broadening in Eq.~\eqref{eq:QbCondition}, thus we introduce a new matching scale $Q_r$ and will present its natural defining condition further ahead. With this separation of the potential, the solution of the Schrödinger Eq.~\eqref{eq:EkernelDef} can be considered order by order as a perturbative series. The NLO correction for the HO solution can be found as the first iteration of a Dyson-like equation~\cite{Barata:2021wuf}, resulting in
\begin{align}\label{eq:KNLO}
    & \cK^{\rm NLO}(t_2,\z;t_1,\y) = -\int_{t_1}^{t_2}ds\int_\x\cK^{\rm LO}(t_2,\z;s,\x)\delta \cV(\x,s) \cK^{\rm LO}(s,\x;t_1,\y)\,.
\end{align}
To consider all NLO corrections to the gluon emission spectrum in Eq.~\eqref{eq:QuarkGluonSpectrum}, one should also take into account the correction associated with the expansion of the broadening kernel that we saw in Section~\ref{subsec:BroadeningIOE}. Thus, schematically, the full answer to NLO accuracy can be rewritten as
\begin{align}
   \frac{d\cI^{\text{LO+NLO}}}{d\o d^2\k} \sim \cP^{\text{LO}}\cK^{\text{LO}}+\cP^{\text{NLO}}\cK^{\text{LO}}+\cP^{\text{LO}}\cK^{\text{NLO}}\,,
\end{align}
where the first term corresponds to the LO solution and the last two are the NLO corrections. The matching scale for the emission kernel is determined by the following transcendental equation~\cite{Barata:2020sav}
\begin{align}\label{eq:QrCondition}
    Q_r^2 = \sqrt{\hat{q}_0\o \log\frac{Q_r^2}{\mu^2_{*}}}\,.
\end{align}
This condition was obtained in~\cite{Barata:2020sav} by imposing the convergence of the IOE series for the gluon energy spectrum in the limit of vanishing gluon energy $\omega\rightarrow 0$.
A detailed analysis of the medium-induced gluon emission spectrum off a quark in the IOE can be found in~\cite{Barata:2021wuf}. The authors find that the IOE approach successfully captures the relevant features of both MS and SH regions by comparing it analytically and numerically with the HO~\cite{Blaizot:2013vha} and GLV~\cite{Gyulassy:2000er} results. As mentioned above, this spectrum can be obtained as the limit $\delta \n\rightarrow0$ of the interference term in the antenna emission spectrum in Eq.~\eqref{eq:InterferenceDef}. Thus, this limit will provide a non-trivial test of our result, which we derive in the following section.
\end{itemize}

\section{Antenna spectrum within the IOE}~\label{sec:AntennaIOE}
\indent In this section, we apply the IOE approach to compute the $\gamma \rightarrow q\bar q$ antenna gluon emission spectrum, as defined in Section \ref{sec:antenna} in Eqs.~\eqref{eq:AntennaSpectrumDef},~\eqref{RqDef} and~\eqref{eq:InterferenceDef}. As shown in the previous Section~\ref{sec:IOE}, this method involves evaluating the dipole potential as a perturbative series around the harmonic oscillator solution, including a small correction that provide a more accurate description of the emission spectrum at higher gluon energies and/or transverse momenta. The main result of this work is the computation of the NLO correction to the interference term of the emission spectrum in Eq.~\eqref{eq:InterferenceDef}. The direct terms at NLO order can be found in~\cite{Barata:2021wuf} or through the limit $\delta\n \rightarrow 0$ of our result. For the case of a static and finite size medium, it is convenient to decompose the time integrals required for the spectrum calculation into three distinct contributions
%
\begin{align}\label{eq:IntegralSplited}
    \int_0^\infty dt_2 \int_0^{t_2}dt_1 =  \int_0^L dt_2 \int_0^{t_2}dt_1+ \int_L^\infty dt_2 \int_0^{L}dt_1+\int_L^\infty dt_2 \int_L^{t_2}dt_1\,,
\end{align}
%
For clarity, the three integration regions are referred to as \textbf{in-in}, \textbf{in-out} and \textbf{out-out}, respectively. In what follows, we consider these three contributions separately, identifying them with the corresponding subscripts, leading to:
\begin{align}\label{eq:InterferenceSeparated}
    \notag\mathcal{J}_{\rm in-in}&=\Re \int_0^{L} dt_2 \int_0^{t_2} dt_1 \left (1-\Delta_{\rm med}(t_1)\right ) 
    \\
    &\hspace{1cm}\times\int_\z e^{-i\bkappa\cdot\z}(\partial_{\y} - i\omega \delta \n) \cdot \partial_{\z} \, \mathcal{P}(L,\z;t_2)\cK(t_2,\z;t_1,\y)\Big|_{\y = \delta \n t_1}\,,
    \\
    \notag\mathcal{J}_{\rm in-out}&=\Re \int_L^{\infty} dt_2 \,e^{-\varepsilon t_2}\int_0^{L} dt_1 \left(1-\Delta_{\rm med}(t_1)\right) 
    \\
    &\hspace{1cm}\times\int_\z e^{-i\bkappa\cdot\z}(\partial_{\y} - i\omega \delta \n) \cdot \partial_{\z}\,\int_{\x}\cK_0(t_2,\z;L,\x)\cK(L,\x;t_1,\y)\Big|_{\y = \delta \n t_1}\,,
    \\
    \notag\mathcal{J}_{\rm out-out}&=\Re \int_L^{\infty} dt_2 \,e^{-\varepsilon t_2} \int_L^{t_2} dt_1 \,e^{-\varepsilon t_1} \left(1-\Delta_{\rm med}(t_1)\right)
    \\ 
    &\hspace{1cm}\times\int_\z e^{-i\bkappa\cdot\z}(\partial_{\y} - i\omega \delta \n) \cdot \partial_{\z}\, \cK_0(t_2,\z;t_1,\y)\Big|_{\y = \delta \n t_1}\,,
\end{align}
where $\cK_0$ is the vacuum propagator Eq.~\eqref{eq:KVac}.\footnote{For the in-out term, we used the following property of the emission kernel: $\cK(t_2,\z;t_1,\y) = \int d^2\x \cK_0(t_2,\z;L,\x)\cK(L,\x;t_1,\y)$, which arises directly.} For the in-out and out-out contribution, we have used the broadening kernel in the vacuum limit, a consequence of imposing $t_2 > L$ in Eq.~\eqref{eq:BKernelDef1}.

Following the IOE prescription, we again consider the expansion of the dipole potential, entering both $\cP$ and $\cK$, keeping in mind that the respective matching scales obey different defining equations, Eq.~\eqref{eq:QbCondition} and Eq.~\eqref{eq:QrCondition}, and must therefore be treated distinctly. Unlike the direct radiation terms in Eq.~\eqref{RqDef}, the interference spectrum also includes the decoherence factor $\Delta_{\rm med}(t)$. Although the dipole potential entering the definition of the decoherence factor in Eq.~\eqref{eq:CoherenceFDef} would, in principle, also require an IOE treatment, we choose to keep it in its full form in Eq.~\eqref{eq:DipolePotentialGW}, since in this case the integral in Eq.~\eqref{eq:CoherenceFDef} can be solved analytically. This adds corrections 
which go beyond NLO, but that are in principle negligible compared to the lowest order contributions. The expansion of $\Delta_{\rm med}(t)$ will be addressed later, in the context of the numerical analysis and phenomenological applications.
Hence, the calculation of the interference spectrum Eq.~\eqref{eq:InterferenceDef} within the IOE approach can be schematically represented according to the order in $\delta\cV$, such that the first non-trivial corrections read
\begin{align}\label{eq:CalculationScheme}
   \cJ^{\rm LO+NLO}\sim \left(\cP^{\text{LO}}\cK^{\text{LO}}+\cP^{\text{NLO}}\cK^{\text{LO}}+\cP^{\text{LO}}\cK^{\text{NLO}}\right)\left (1-\Delta_{\rm med}\right)^{\text{Full}}\,.
\end{align}
 The results in the following subsections will be presented by splitting each contributions according to the three time integration regions defined in Eq.~\eqref{eq:InterferenceSeparated}.

\subsection{LO interference contribution}
\vspace*{-3mm}~\\
\indent We begin with a warm-up calculation of the LO result, which was obtained in~\cite{Mehtar-Tani:2012mfa, Casalderrey-Solana:2011ule}. The LO broadening kernel is only different from unity for the \textbf{in-in} contribution, where it reads
\begin{align}
    \cP^{\rm LO}(L, \z; t_2) = e^{-\frac{1}{4}Q_s^2(L,t_2)\z^2}\,,
\end{align}
with $Q_s^2(L,t_2) = \hat q_0 (L-t_2)\log\frac{Q_b^2}{\mu_{\ast}^2}$. The LO emission kernel $\cK$ is given by Eq~\eqref{eq:KLO} with the replacement $\hat q_0 \rightarrow \hat q_0\log\frac{Q_r^2}{\mu_{\ast}^2}$. Then the \textbf{in-in} contribution reads 
\begin{align}
	& \mathcal{J}^{\rm LO}_{\rm in-in}  = \Re \int_0^L dt_2 \int_0^{t_2} dt_1 \left(1-\Delta_{\rm med}(t_1)\right)e^{i\frac{\omega}{2}\delta\n^2 t_1}\nn
	& \hspace{0.5cm}\times\int_{\z} e^{-i\bkappa\cdot\z} \mathcal{P}^{\rm LO}(L,\z; t_2)(\partial_{\y}-i\omega\delta \n)\cdot\partial_{\z}\left.\cK^{\rm LO}(t_2,\z;t_1,\y)\right|_{\y = \delta\n t_1} + \rm{sym.} \, .
\end{align}
The integrand can be directly simplified in this approximation
\begin{align}\label{eq:derivatives_in_in}
	 \notag & (\partial_{\y}-i\omega\delta \n)\cdot\partial_{\z}\left.\cK^{\rm LO}(t_2,\z;t_1,\y)\right|_{\y = \delta\n t_1} =
     \\
	 \notag&  \hspace{1cm}\frac{\omega^2}{2\pi S_{21}^3}\Big[i\omega \delta\n\cdot\z\Big(t_1-C_{21}(-C_{21}t_1+S_{21})\Big) + i\omega\delta\n^2 t_1(-C_{21}t_1+S_{21})
     \\
     &\hspace{3cm}-2S_{21} -i\omega C_{21}\z^2\Big]\Exp{\frac{i\omega}{2S_{21}}\Big[C_{21}(\z^2+t_1^2\delta\n^2)-2t_1\delta\n\cdot\z\Big]} \, ,
\end{align}
such that we obtain
\begin{align}
	\notag& \cJ^{\rm LO}_{\rm  in-in} = \frac{\omega^3}{2\pi}\Re\int_0^L dt_2 \int_0^{t_2}dt_1(1-\Delta_{\rm med}(t_1))e^{i\frac{\omega}{2}\delta\n^2 t_1(1+\frac{t_1}{T_{21}})}\frac{1}{S_{21}^3}
    \\
    \notag &\hspace{0.5cm}\times \int_\z e^{-i\q_{21}\cdot\z}e^{-\hat P_{21}^2\z^2/4}\Big[-2S_{21} + i\omega\delta\n^2 t_1(-C_{21}t_1+S_{21})
    \\
    &\hspace{2.5cm}+i\omega \delta\n\cdot\z\Big(t_1-C_{21}(-C_{21}t_1+S_{21})\Big)-i\omega C_{21}\z^2 \Big] + \rm sym.\,,
\end{align}
where we have defined $\q_{21} = \bkappa + \frac{\omega t_1}{S_{21}}\delta \n$, $T_{21} = S_{21}/C_{21}$,  $\hat P_{21}^2 = Q_s^2(L,t_2) -\frac{2i\omega}{T_{21}}$.
The $\z$ integrals are simple exponential integrals with polynomials, so after carrying them out the \textbf{in-in} contribution reads
\begin{align}\label{eq:in_in_LO}
    \notag  &\cJ^{\rm LO}_{\rm  in-in} = 2\omega^2\Re\int_0^L dt_2 \int_0^{t_2}dt_1(1-\Delta_{\rm med}(t_1))e^{i\frac{\omega}{2}\delta\n^2 t_1(1+\frac{t_1}{T_{21}})}
    \\
    \notag &\hspace{1.5cm} \times \frac{e^{-\q_{21}^2/\hat P_{21}^2}}{\hat P_{21}^2 S_{21}^3}\Bigg[-2S_{21} + i\omega\delta\n^2 t_1(-C_{21}t_1+S_{21})-\frac{4i\omega C_{21}}{\hat P_{21}^2}\Big(1 - \frac{\q_{21}^2}{\hat P_{21}^2}\Big)
    \\
    & \hspace{5cm}+2\omega \frac{\delta\n\cdot\q_{21}}{\hat P_{21}^2}\Big(t_1-C_{21}(-C_{21}t_1+S_{21})\Big) \Bigg] + \rm sym. \, .
\end{align}
The \textbf{in-out} contribution, which reads
\begin{align}
	& \cJ^{\rm LO}_{\rm  in-out}  = \Re \int_L^{\infty} dt_2 \,e^{-\varepsilon t_2}\int_0^L dt_1 \left(1-\Delta_{\rm med}(t_1)\right)e^{i\frac{\omega}{2}\delta\n^2 t_1}\nn
	& \hspace{0.5cm}\times\int_{\z} e^{-i\bkappa\cdot\z}\int_\x\partial_{\z}\cK_0(t_2,\z;L,\x)\cdot (\partial_{\y}-i\omega\delta \n)\left.\cK^{\rm LO}(L,\x;t_1,\y)\right|_{\y = \delta\n t_1} + \rm{sym.} \, ,
\end{align}
can also be simplified since the $\z$ and $t_2$ integrals can be performed:
\begin{align}\label{eq:CoolIntegral}
	\int_L^{\infty} dt_2 \,e^{-\varepsilon t_2} \int_\z e^{-i\bkappa\cdot\z}\partial_\z\cK_0(t_2,\z;L,\x) = 2\omega e^{-i\bkappa\cdot\x}\frac{\bkappa}{\bkappa^2}\,.
\end{align}
This leads to the compact expression for the spectrum
\begin{align}\label{eq:in_out_LO}
	& \mathcal{J}^{\rm LO}_{\rm in-out}  = -\frac{2\omega}{\bkappa^2}\Re\,i\int_0^L dt_1 \left(1-\Delta_{\rm med}(t_1)\right)e^{i\frac{\omega}{2}\delta\n^2 t_1(1+\frac{t_1}{T_{L1}})}\nn
	& \hspace{0.5cm}\times \frac{e^{-\q_{L1}^2/\hat P_{L1}^2}}{C_{L1}^2}\bkappa\cdot\Big(\q_{L1} + \frac{i\hat P_{L1}^2}{2}\delta\n(-C_{L1}t_1 + S_{L1})\Big)+ \rm{sym.}\,,
\end{align}
with $\q_{L1} = \bkappa + \frac{\omega t_1}{S_{L1}}\delta\n$ and $\hat P_{L1}^2 = -\frac{2i\omega}{T_{L1}}$. 

The last LO contribution is the \textbf{out-out} term, which corresponds to radiation occurring entirely outside the medium. In this case, both kernels take their vacuum forms and do not require expansion within the IOE framework. Therefore, 
because we calculate the decoherence factor $\Delta_{\rm med}$ with the full potential in Eq.~\eqref{eq:DipolePotentialGW}, this contribution does not receive what we define to be NLO corrections
and it is given by
\begin{align}
	& \mathcal{J}_{\rm out-out}  = \left(1-\Delta_{\rm med}(L)\right)\Re \int_L^{\infty} dt_2 \,e^{-\varepsilon t_2}\int_L^{t_2} dt_1\,e^{-\varepsilon t_1} e^{i\frac{\omega}{2}\delta\n^2 t_1}\nn
	& \hspace{0.5cm}\times\int_{\z} e^{-i\bkappa\cdot\z} (\partial_{\y}-i\omega\delta \n)\cdot\partial_{\z}\left.\cK_0(t_2,\z;t_1,\y)\right|_{\y = \delta\n t_1} + \rm{sym.}
\end{align}
After performing all derivatives and integrals, and summing the $\text{sym}$ term, we obtain

\begin{align}\label{eq:Ioutout} 
    \cJ_{\rm out-out} = 4\omega^2\frac{\bkappa\cdot\bar\bkappa} {\bkappa^2\bar\bkappa^2}\Big(1 - \Delta_{\rm med}(L)\Big)\cos\left(\frac{\bkappa^2 - \bar\bkappa^2}{2\omega} L\right)\,,
\end{align}

which one can compare directly with e.g. \cite{Mehtar-Tani:2012mfa}.

\subsection{NLO interference contribution}
\vspace*{-3mm}~\\
\indent As schematically illustrated in Eq.~\eqref{eq:CalculationScheme}, both the broadening and emission kernels must be expanded according to the IOE prescription, each with its respective matching scale, $Q_b$ and $Q_r$. This leads to two distinct NLO contributions. For reference, the NLO corrections to the kernels are given by
\begin{align} 
    &\cP^{\rm NLO}(L,\z;t_2) = \delta \cV(\z)\,\cP^{\rm LO}(L,\z;t_2)\,,
\end{align}
for the broadening kernel, and
\begin{align} 
    &\cK^{\rm NLO}(t_2,\z;t_1,\y) = -\int_{t_1}^{t_2} ds \int_\x \cK^{\rm LO}(t_2,\z;s,\x)\, \delta \cV(\x,s)\, \cK^{\rm LO}(s,\x;t_1,\y)\,,
\end{align}
for the emission kernel. To distinguish between the two types of corrections, we will denote the NLO terms with additional superscripts b and r corresponding to the terms containing only the broadening expansion or only emission kernel expansion, respectively. 

\paragraph{In-In Broadening\\}
The first \textbf{in-in} contribution, associated with the broadening kernel expansion, reads
\begin{align}
	& \cJ_{\rm in-in}^{\rm NLO,\, b} = -\Re\int_0^L dt_2 \int_0^{t_2}dt_1(1-\Delta_{\rm med}(t_1))e^{i\frac{\omega}{2}\delta\n^2 t_1}\nn
	& \hspace{0.5cm}\times\int_\z e^{-i\bkappa\cdot\z}\cP^{\rm LO}(L,\z;t_2)\delta \cV(\z)(\partial_{\y}-i\omega\delta \n)\cdot\partial_{\z}\left.\cK^{\rm LO}(t_2,\z;t_1,\y)\right|_{\y = \delta\n t_1} + \rm sym. \, .
\end{align}
After explicitly applying the derivatives to the LO emission kernel as in Eq.~\eqref{eq:derivatives_in_in}, this contribution becomes
\begin{align}
    \notag& \cJ_{\rm in-in}^{\rm NLO,\, b} = -\frac{\hat q_0\omega^2}{8\pi}\Re\int_0^L dt_2 \int_0^{t_2}dt_1(1-\Delta_{\rm med}(t_1))e^{i\frac{\omega}{2}\delta\n^2 t_1(1+\frac{t_1}{T_{21}})}
    \\
    \notag &\hspace{1cm}\times\frac{(L-t_2)}{S_{21}^3}\int_\z e^{-i\q_{21}\cdot\z}e^{-P_{21}^2 \z^2/4}\z^2\log\frac{1}{Q_b^2\z^2}
    \\
    \notag &\hspace{2cm}\times\Big[-2S_{21} + i\omega\delta\n^2 t_1(-C_{21}t_1+S_{21})-i\omega C_{21}\z^2
    \\
    &\hspace{5cm}+i\omega \delta\n\cdot\z\Big(t_1-C_{21}(-C_{21}t_1+S_{21})\Big) \Big]+ \rm sym.\,,
\end{align}
where one should recall the definitions $\q_{21} = \bkappa + \frac{\omega t_1}{S_{21}}\delta \n$ and $\hat P_{21}^2 = Q_s^2(L,t_2) -\frac{2i\omega}{T_{21}}$. In contrast to the LO case, now the $\z$ integrals have a more involved form:
\begin{align}
	&  \int_\z e^{-i\q_{21}\cdot\z}e^{- \hat{P}_{21}^2 \z^2/4} \z^2\log\frac{1}{Q^2\z^2} = \frac{2\pi}{\q_{21}^4} I_a^{(3)}\left(\frac{\q_{21}^2}{\hat P_{21}^2},\frac{\q_{21}^2}{Q_b^2}\right)\nn
	& \int_\z e^{-i\q_{21}\cdot\z}e^{- \hat{P}_{21}^2 \z^2/4} \z^3\log\frac{1}{Q^2\z^2}   = -2\pi i \frac{\q_{21}}{\q_{21}^6} I_b^{(4)}\left(\frac{\q_{21}^2}{\hat P_{21}^2},\frac{\q_{21}^2}{Q_b^2}\right)\nn
	& \int_\z e^{-i\q_{21}\cdot\z}e^{- \hat{P}_{21}^2 \z^2/4} \z^4\log\frac{1}{Q^2\z^2}  = \frac{2\pi}{\q_{21}^6} I_a^{(5)}\left(\frac{\q_{21}^2}{\hat P_{21}^2},\frac{\q_{21}^2}{Q_b^2}\right)\,,
\end{align}
where the shorthand notations for the integrals read
\begin{align}\label{eq:MoilereIntegrals}
	& I^{(n)}_a(x,y) = \int_0^{\infty} dz\, z^n J_0(z) \log\frac{y}{z^2}e^{-\frac{z^2}{4x}} \,,\nn
	& I^{(n)}_b(x,y) = \int_0^{\infty} dz\, z^n J_1(z) \log\frac{y}{z^2}e^{-\frac{z^2}{4x}}\,.
\end{align} 
Despite their complicated form, these integrals can be evaluated analytically~\cite{Barata:2021wuf} and we present their explicit form in Section~\ref{sec:FinalResults}. Combining all these results, we obtain the final expression for this contribution
\begin{align}\label{eq:in_in_b}
	& \cJ_{\rm in-in}^{\rm NLO,\, b} = -\frac{\hat q_0\omega^2}{4}\Re\int_0^L dt_2 \int_0^{t_2}dt_1(1-\Delta_{\rm med}(t_1))e^{i\frac{\omega}{2}\delta\n^2 t_1(1+\frac{t_1}{T_{21}})}\frac{(L-t_2)}{\q_{21}^4S_{21}^3}\nn
	\notag& \hspace{0.5cm}\times\Bigg[\Big(-2S_{21} + i\omega\delta\n^2 t_1(-C_{21}t_1+S_{21})\Big)I_a^{(3)}\left(\frac{\q_{21}^2}{\hat P_{21}^2},\frac{\q_{21}^2}{Q_b^2}\right)-i\frac{\omega C_{21}}{\q_{21}^2}I_a^{(5)}\left(\frac{\q_{21}^2}{\hat P_{21}^2},\frac{\q_{21}^2}{Q_b^2}\right)\\
	&  \hspace{2cm}+\omega \frac{\q_{21}\cdot\delta\n}{\q_{21}^2}\Big(t_1-C_{21}(-C_{21}t_1+S_{21})\Big)I_b^{(4)}\left(\frac{\q_{21}^2}{\hat P_{21}^2},\frac{\q_{21}^2}{Q_b^2}\right) \Bigg] + \rm sym. \, .
\end{align}

\paragraph{In-In Radiative\\}
The next \textbf{in-in} contribution is associated with the expansion of the emission kernel and it reads
\begin{align}
    \notag& \cJ_{\rm in-in}^{\rm NLO,\, r}  = -\Re\int_0^L dt_2 \int_0^{t_2}dt_1(1-\Delta_{\rm med}(t_1))e^{i\frac{\omega}{2}\delta\n^2 t_1} 
    \\
    \notag&\hspace{0.5cm}\times\int_\x \delta \cV(\x)\int_{t_1}^{t_2}ds\,(\partial_{\y}-i\omega\delta \n) \cK^{\rm LO}(s,\x;t_1,\y)\Big|_{\y = \delta\n t_1}
    \\
    &\hspace{2.5cm}\cdot\int_\z \partial_{\z}\cK^{\rm LO}(t_2,\z;s,\x)e^{-i\bkappa\cdot\z}\cP^{\rm LO}(L,\z;t_2)+ \rm sym.
\end{align}
Applying the derivatives following Eq.~\eqref{eq:derivatives_in_in} and explicitly integrating over $\z$ this contribution reduces to
\begin{align}
    \notag& \cJ_{\rm in-in}^{\rm NLO,\, r}  = -\frac{\hat q_0\omega^4}{4\pi}\Re\int_0^L dt_2 \int_0^{t_2}dt_1(1-\Delta_{\rm med}(t_1))\int_{t_1}^{t_2}ds\,e^{i\frac{\omega}{2}\delta\n^2 t_1(1+\frac{t_1}{T_{s1}})}
    \\
    \notag&\hspace{0.5cm}\times\frac{e^{-\bkappa^2/\hat P_{2s}^2}}{\hat P_{2s}^4S_{2s}^2S_{s1}^2}\int_\x e^{-i\q_{2s1}\cdot\x} e^{-\frac{1}{4}G_{2s1} \x^2}\x^2\log\frac{1}{Q_r^2\x^2} \,\Bigg[2iC_{2s}(-C_{s1}t_1 + S_{s1})\bkappa\cdot\delta\n
    \\
    &\hspace{2cm}+\Big(2iC_{2s}\bkappa + Q_s^2(L,t_2)(-C_{s1}t_1 + S_{s1})\delta\n\Big)\cdot\x + Q_s^2(L,t_2)\x^2\Bigg]+ \rm sym.\,,
\end{align}
where $\hat P_{2s}^2 = Q_s^2(L,t_2) - \frac{2i\omega}{T_{2s}}$, $\q_{2s1} = -\frac{2i\omega}{\hat P_{2s}^2S_{2s}}\bkappa + \frac{\omega t_1}{S_{s1}}\delta \n$ and $G_{2s1} = \frac{4\omega^2}{S_{2s}^2\hat P_{2s}^2}-2i\omega\left(\frac{1}{T_{2s}}+\frac{1}{T_{s1}}\right)$. After rewriting this expression in terms of the integrals defined in Eq.~\eqref{eq:MoilereIntegrals}, we obtain the final expression for this contribution
\begin{align}\label{eq:in_in_NLO_r}
	&  \cJ_{\rm in-in}^{\rm NLO,\, r}  = -\frac{\hat q_0\omega^4}{2}\Re\int_0^L dt_2 \int_0^{t_2}dt_1(1-\Delta_{\rm med}(t_1))\int_{t_1}^{t_2}ds\,e^{i\frac{\omega}{2}\delta\n^2 t_1(1+\frac{t_1}{T_{s1}})}\frac{e^{-\bkappa^2/\hat P_{2s}^2}}{\hat P_{2s}^4S_{2s}^2S_{s1}^2\q_{2s1}^4}\nn
	& \hspace{0.5cm}\times\Bigg[2iC_{2s}(-C_{s1}t_1 + S_{s1})\bkappa\cdot\delta\n\, I_a^{(3)}\left(\frac{\q_{2s1}^2}{G_{2s1}},\frac{\q_{2s1}^2}{Q_r^2}\right)+\frac{Q_s^2(L,t_2)}{\q_{2s1}^2}I_a^{(5)}\left(\frac{\q_{2s1}^2}{G_{2s1}},\frac{\q_{2s1}^2}{Q_r^2}\right) \nn
	& \hspace{1cm}-i (2iC_{2s}\bkappa + Q_s^2(L,t_2)(-C_{s1}t_1 + S_{s1})\delta\n)\cdot \frac{\q_{2s1}}{\q_{2s1}^2}I_b^{(4)}\left(\frac{\q_{2s1}^2}{G_{2s1}},\frac{\q_{2s1}^2}{Q_r^2}\right)\Bigg]+ \rm sym. \, .
\end{align}

\paragraph{In-Out Radiative\\}
We now turn to the \textbf{in-out} NLO contributions. Since this part of the spectrum does not explicitly contain $\cP$ (see the definition in Eq.~\eqref{eq:InterferenceSeparated}), it receives only the IOE correction associated with the expansion of the emission kernel, which reads
\begin{align}
    \notag&\cJ_{\rm in-out}^{\rm NLO,\, r}  = -\Re \int_0^{L}dt_1(1-\Delta_{\rm med}(t_1))\int_{t_1}^{L}ds\,e^{i\frac{\omega}{2}\delta\n^2 t_1}
    \\
    \notag &\hspace{0.5cm}\times\int_\x \delta \cV(\x)\, (\partial_{\y}-i\omega\delta \n)\left.\cK^{\rm LO}(s,\x;t_1,\y)\right|_{\y = \delta\n t_1}
    \\
    &\hspace{2cm}\cdot\int_{\u} \cK^{\rm LO}(L,\u;s,\x)\int_L^{\infty} dt_2\,e^{-\varepsilon t_2}\int_{\z} e^{-i\bkappa\cdot\z}\partial_{\z}\cK_0(t_2,\z;L,\u) + \rm sym.\, .
\end{align}
We start by using the result in Eq.~\eqref{eq:CoolIntegral} for the $\z$ and $t_2$ integrals involving the vacuum emission kernel and then perform the integral over $\u$
\begin{align}
	\int_\u e^{-i\bkappa\cdot\u}\cK^{\rm LO}(L,\u;s,\x) = \frac{1}{C_{Ls}}e^{-i\frac{T_{Ls}}{2\omega}\bkappa^2}e^{-i\bkappa\cdot\x/C_{Ls}}e^{-i\frac{\omega T_{Ls}}{2}\Omega^2\x^2}\,.
\end{align}
Taking the remaining derivatives as done above, we find
\begin{align}
    \notag& \cJ_{\rm in-out}^{\rm NLO,\, r}  = \frac{\hat{q}_0\omega^3}{4\pi\bkappa^2}\Re\int_0^{L}dt_1 (1-\Delta_{\rm med}(t_1))\int_{t_1}^L ds\,e^{i\frac{\omega}{2}\delta\n^2 t_1(1+\frac{t_1}{T_{s1}})}
    \\
    \notag& \hspace{1cm}\times\frac{e^{-i\frac{T_{Ls}}{2\omega}\bkappa^2}}{C_{Ls}S_{s1}^2}\int_\x e^{-i\q_{Ls1}\cdot\x}e^{-G_{Ls1}\x^2/4}\x^2\log\frac{1}{Q_r^2\x^2}
    \\
    &\hspace{2cm}\times\Big[\bkappa\cdot\x + \bkappa\cdot\delta\n(-C_{s1}t_1 + S_{s1})\Big]+ \rm sym.\,,
\end{align}
where $\q_{Ls1} = \frac{1}{C_{Ls}}\bkappa + \frac{\omega t_1}{S_{s1}} \delta\n$ and $G_{Ls1} = 2i\omega T_{Ls} \Omega^2 - \frac{2i\omega}{T_{s1}}$. Using the shorthand notation for the integrals defined in Eq.~\eqref{eq:MoilereIntegrals}, we can rewrite this NLO contribution as
\begin{align}\label{eq:in_out_NLO_r}
	& \cJ_{\rm in-out}^{\rm NLO,\, r}  = \frac{\hat q_0\omega^3}{2\bkappa^2}\Re\int_0^{L}dt_1(1-\Delta_{\rm med}(t_1))\int_{t_1}^L ds\,e^{i\frac{\omega}{2}\delta\n^2 t_1(1+\frac{t_1}{T_{s1}})}\nn
	\notag & \hspace{1cm}\times \frac{e^{-i\frac{T_{Ls}}{2\omega}\bkappa^2}}{C_{Ls}S_{s1}^2\q_{Ls1}^4}\Bigg[-i\frac{\bkappa\cdot\q_{Ls1}}{\q_{Ls1}^2}I_b^{(4)}\left(\frac{\q_{Ls1}^2}{G_{Ls1}},\frac{\q_{Ls1}^2}{Q_r^2}\right)
    \\
    & \hspace{4cm}+ \bkappa\cdot\delta\n(-C_{s1}t_1 + S_{s1})I_a^{(3)}\left(\frac{\q_{Ls1}^2}{G_{Ls1}},\frac{\q_{Ls1}^2}{Q_r^2}\right)\Bigg]+ \rm sym. \, .
\end{align}

As noted in Eq.~\eqref{eq:Ioutout}, the \textbf{out-out} contribution corresponds to the emission occurring outside the medium in both amplitude and complex conjugate amplitude, thus it does not involve IOE corrections at any order in either the emission or broadening kernels.

\subsection{Vanishing dipole size limit}\label{subsec:AsymptoticBehavior}
\vspace*{-3mm}~\\
\indent As previously discussed, in the vanishing dipole size limit, $\delta\n \rightarrow 0$, the interference term fully cancels the direct contributions, such that $\cJ\big|_{\delta \n \rightarrow 0} = \cR_{q}\big|_{\delta\n\rightarrow 0} = \cR_{\bar q}\big|_{\delta\n\rightarrow 0}\equiv \cR$ and radiation cannot occur due to the pair being resolved as a color singlet state.\footnote{From now on, we use $\cR$ to represent any of the two direct contributions in this limit.} In this limit, $\cR$ can be understood as the gluon emission off a single quark (see Eq.~\eqref{eq:QuarkGluonSpectrum}), providing a non-trivial test of our calculation of the IOE corrections to the interference term, since $\cR$ has already been studied in detail in~\cite{Barata:2021wuf}.
Starting with the LO terms, the \textbf{in-in} contribution in Eq.~\eqref{eq:in_in_LO} in this limit takes the form
\begin{align}
	&  \cR_{\rm in-in}^{\rm LO} = 4\omega^2\Re\int_0^L dt_2 \int_0^{t_2}dt_1\frac{e^{-\k^2/\hat P_{21}^2}}{\hat P_{21}^2 S_{21}^3} \left[-2S_{21} -\frac{4i\omega C_{21}}{\hat P_{21}^2}\left(1 - \frac{\k^2}{\hat P_{21}^2}\right) \right]\,,
\end{align}
where now $\bkappa = \bar \bkappa = \k$, so the symmetric contribution is identical to the first term written explicitly. Moreover, one of the time integrals can be evaluated by making the following observation:
\begin{align}
	\frac{e^{-\k^2/\hat P_{21}^2}}{\hat P_{21}^2 S_{21}^3}\left[-2S_{21} -\frac{4i\omega C_{21}}{\hat P_{21}^2}\left(1 - \frac{\k^2}{\hat P_{21}^2}\right)\right] = - 2 \, \partial_{t_1}\left(\frac{e^{-\k^2/\hat P_{21}^2}}{T_{21}\hat P_{21}^2}\right)\,.
\end{align}
Substituting this result back in we obtain
\begin{align}
	&  \cR_{\rm in-in}^{\rm LO} = 8\omega^2\Re\int_0^L dt_2\frac{e^{-\k^2/\hat P_{20}^2}}{T_{20}\hat{P}_{20}^2}\,,
\end{align}
with $\hat{P}_{20}^2 = Q_s^2(L, t_2) - \frac{2i\omega}{T_{20}}$ and $T_{20} = \tan\left(\Omega L\right)/L$. After multiplying by $\alpha_s C_F / \omega^2$, this is in exact agreement with the result found for this contribution in e.g.~\cite{Isaksen:2023nlr}. Turning to the \textbf{in-out} contribution in Eq.~\eqref{eq:in_out_LO}, in this limit it reads
\begin{align}
	& \cR_{\rm in-out}^{\rm LO} = -4\omega \Re\,i\int_0^L dt_1 e^{-\k^2/\hat P_{L1}^2}\frac{1}{C_{L1}^2} = \frac{8\omega^2}{\k^2}\Re\left(e^{-\k^2/\hat P_{L0}^2}-1\right)\,,
\end{align}
which is also agrees with previous results in the literature, after proper normalization. The \textbf{out-out} term in this limit obtains a particularly simple form:
\begin{align} 
     \cR_{\rm out-out} = \frac{4\omega^2}{\k^2}\,, 
\end{align}
which is the standard result in vacuum for gluon emission off a quark, giving rise to both soft and collinear divergences.

We now turn to the test of our NLO results, starting with the two \textbf{in-in} contributions. In the limit of vanishing dipole size, the NLO correction to broadening reads
\begin{align}
    \notag &\cR_{\rm in-in}^{\rm NLO,\,b}= \frac{\hat q_0\omega^2}{\k^4}\Re \int_0^L dt_2\,(L-t_2)
    \\
    &\hspace{1cm}\times\int_0^{t_2}dt_1\frac{1}{S_{21}^2}\left[I_a^{(3)}\left(\frac{\k^2}{\hat P_{21}^2},\frac{\k^2}{Q_b^2}\right) +\frac{i\omega}{2T_{21}\k^2}I_a^{(5)}\left(\frac{\k^2}{\hat P_{21}^2},\frac{\k^2}{Q_b^2}\right)\right] \,,
\end{align}
Using the identity
\begin{align}
    \partial_{t_1}\left(\frac{I_a^{(3)}\left(\frac{\k^2}{\hat P_{21}^2},\frac{\k^2}{Q_b^2}\right)}{T_{21}}\right) = \frac{1}{S_{21}^2}\left[I_a^{(3)}\left(\frac{\k^2}{\hat P_{21}^2},\frac{\k^2}{Q_b^2}\right) +\frac{i\omega}{2T_{21}\k^2}I_a^{(5)}\left(\frac{\k^2}{\hat P_{21}^2},\frac{\k^2}{Q_b^2}\right)\right]\, ,
\end{align}
we obtain the first piece of the NLO spectrum in the $\delta\n\rightarrow 0$ limit:\footnote{After evaluating the $t_1$ integral, another term of the form $I_a^{(3)}\left(\frac{\k^2}{\hat P_{21}^2},\frac{\k^2}{Q_b^2}\right)/T_{21}\big|_{t_1 \rightarrow t_2}$ should appear. At first glance, its structure may seem confusing, as its individual components $1/T_{22}$ and $\hat P_{22}^2$ are divergent. However, after proper treatment of the $\z$ integration inside $I_a^{(3)}$, this contribution vanishes. The same happens for all remaining NLO contributions in the limit we consider.}
\begin{align}
    & \cR_{\rm in-in}^{\rm NLO,\,b}= -\frac{\hat q_0\omega^2}{\k^4}\Re\int_0^L dt_2\frac{L-t_2}{T_{20}}I_a^{(3)}\left(\frac{\k^2}{\hat P_{20}^2},\frac{\k^2}{Q_b^2}\right)\,,
\end{align}
reproducing the result found in~\cite{Barata:2021wuf}. Turning to the second part of the \textbf{in-in} NLO contribution, we find
\begin{align}
	& \cR_{\rm in-in}^{\rm NLO,\,r} = -\hat q_0\omega^4\Re\int_0^L dt_2 \int_0^{t_2}dt_1\int_{t_1}^{t_2}ds\,\frac{e^{-\k^2/\hat P_{2s}^2}}{\hat P_{2s}^4S_{2s}^2S_{s1}^2\q_{2s1}^4}\nn
	& \hspace{1cm}\times\left[iC_{2s}\frac{\hat P_{2s}^2S_{2s}}{\omega}I_b^{(4)}\left(\frac{\q_{2s1}^2}{G_{2s1}},\frac{\q_{2s1}^2}{Q_r^2}\right)+\frac{Q_s^2(L,t_2)}{\q_{2s1}^2}I_a^{(5)}\left(\frac{\q_{2s1}^2}{G_{2s1}},\frac{\q_{2s1}^2}{Q_r^2}\right)\right]\,,
\end{align}
with $\q_{2s1}^2 = -\left(\frac{2\omega}{\hat P_{2s}^2 S_{2s}}\right)^2\k^2$ and $G_{2s1} = \frac{4\omega^2}{S_{2s}^2\hat P_{2s}^2}-2i\omega\left(\frac{1}{T_{2s}}+\frac{1}{T_{s1}}\right)$. In order to make the comparison more explicit, we change the integration limits 
\begin{align}
	\int_0^{t_2} dt_1\int_{t_1}^{t_2} ds = 	\int_0^{t_2} ds\int_{0}^{s} dt_1\,.
\end{align}
In this case, the function can once again be treated as a total derivative, allowing for an explicit evaluation of the $t_1$ integral. This contribution then reads
\begin{align}
	\notag& \cR_{\rm in-in}^{\rm NLO,\,r}= \frac{\hat q_0 \omega}{2\k^4}\Re i\int_0^L dt_2 \int_0^{t_2}ds\,\left(\frac{\hat P_{2s}^2S_{2s}}{-2i\omega}\right)^2 e^{-\k^2/\hat P_{2s}^2}
    \\
    &\hspace{1cm}\times\left[Q_s^2(L,t_2)I_a^{(3)}\left(\frac{\q_{2s1}^2}{G_{2s0}},\frac{\q_{2s1}^2}{Q_r^2}\right)+2C_{2s}\left(\frac{-2i\omega}{\hat P_{2s}^2S_{2s}}\right)\k^2I_b^{(2)}\left(\frac{\q_{2s1}^2}{G_{2s0}},\frac{\q_{2s1}^2}{Q_r^2}\right)\right]\,.
\end{align}
This result, with the appropriate normalization, again reproduces the same contribution found in ~\cite{Barata:2021wuf}. Finally, the last IOE correction is encoded in the \textbf{in-out} contribution
\begin{align}
	& \cR_{\rm in-out}^{\rm NLO,\,r} = -\frac{\hat q_0\omega^3}{\k^6}\Re i\int_0^{L}dt_1\int_{t_1}^L ds\frac{C_{Ls}^4}{S_{s1}^2}  e^{-i\frac{T_{Ls}}{2\omega}\k^2}I_b^{(4)}\left(\frac{\q_{Ls1}^2}{G_{Ls1}},\frac{\q_{Ls1}^2}{Q_r^2}\right)\,,
\end{align}
with $\q_{Ls1} = \frac{\k}{C_{Ls}}$ and $G_{Ls1} =2i\omega T_{Ls}\Omega^2 - \frac{2i\omega}{T_{s1}}$,
which after integration over $t_1$ leads to 
\begin{align}
    &\cR_{\rm in-out}^{\rm NLO,\,r} = \frac{2\hat q_0\omega^2}{\k^4}\Re\int_0^{L}ds\, C_{Ls}^2e^{-i\frac{T_{Ls}}{2\omega}\k^2}I_b^{(4)}\left(\frac{\k^2}{C_{Ls}^2G_{Ls0}},\frac{\k^2}{C_{Ls}^2Q_r^2}\right)\,.
\end{align}
%
in agreement with~\cite{Barata:2021wuf}.

\section{Summary of key expressions}~\label{sec:FinalResults}
\indent Before turning to the numerical analysis of the antenna spectrum, we summarize here the results obtained in the previous section, along with the notations used throughout. In our calculations, we focused on the case of a static, finite medium, which is characterized by the following functions:
\begin{align}
    \nonumber& S_{21} = \frac{\sin ((t_2-t_1)\Omega)}{\Omega}\,,\quad C_{21} = \cos ((t_2-t_1)\Omega)\,, \quad T_{21} = \frac{S_{21}}{C_{21}}\,,\quad \Omega = \frac{1-i}{2}\sqrt{\frac{\hat q_0}{\omega}\log\frac{Q_r^2}{\mu^2_{*}}}\,.
\end{align}
Using the IOE prescription, we defined the matching scales $Q_b$ and $Q_r$, which, together with the saturation scale $Q_s$, satisfy the following relations:
\begin{align}
    \nonumber& Q_s^2(t_2,t_1)= \hat q_0 \log\frac{Q_b^2}{\mu_*^2} (t_2-t_1)\,,\quad  Q_b^2 =\hat q_0  \log\frac{Q_b^2}{\mu_*^2} \,L\,,\quad Q_r^2 = \sqrt{\hat q_0 \omega \log\frac{Q_r^2}{\mu_*^2}}\,.
\end{align}
During the calculation we have also defined the following list of transverse momenta 
\begin{align}
    &\q_{21} = \bkappa + \frac{\omega t_1}{S_{21}}\delta \n\,,\quad \q_{L1} = \bkappa + \frac{\omega t_1}{S_{L1}}\delta\n\,,\nn
    \notag&\q_{2s1} = -\frac{2i\omega}{\hat P_{2s}^2S_{2s}}\bkappa + \frac{\omega t_1}{S_{s1}}\delta \n\,,\quad \q_{Ls1} = \frac{1}{C_{Ls}}\bkappa+\frac{\omega t_1}{S_{s1}}\delta\n\,,
\end{align}
and the characteristic phases
\begin{align}
    &  \hat P_{21}^2 = Q_s^2(L,t_2) -\frac{2i\omega}{T_{21}}\,,\quad\hat P_{2s}^2 = Q_s^2(L,t_2) - \frac{2i\omega}{T_{2s}}\,,\quad\hat P_{L1}^2 = -\frac{2i\omega}{T_{L1}}\,,\nn
	\notag&  G_{2s1} = \frac{4\omega^2}{S_{2s}^2\hat P_{2s}^2}-2i\omega\left(\frac{1}{T_{2s}}+\frac{1}{T_{s1}}\right)\,,\quad\quad G_{Ls1} =2i\omega T_{Ls}\Omega^2 - \frac{2i\omega}{T_{s1}}\,.
\end{align}
We have also introduced the two types of integral families
\begin{align}
    & I^{(n)}_a(x,y) = \int_0^{\infty} dz\, z^n J_0(z) \log\frac{y}{z^2}e^{-\frac{z^2}{4x}} \nn
	\notag & I^{(n)}_b(x,y) = \int_0^{\infty} dz\, z^n J_1(z) \log\frac{y}{z^2}e^{-\frac{z^2}{4x}}\,.
\end{align}
They can be evaluated analytically, however, in this calculation we only need a particular set of integrals presented above. Thus, we use the expressions for $I^{(3)}_a$ and $I^{(2)}_b$ introduced in \cite{Barata:2021wuf}, which, together with the observations $I^{(5)}_a = 4 x^2 \frac{\partial}{\partial x} I^{(3)}_a$ and $I^{(4)}_b = 4 x^2 \frac{\partial}{\partial x} I^{(2)}_b$, provide the list of integrals entering the spectrum:
\begin{align}
    \notag& I^{(3)}_a(x,y) = 8x^2\left[1 -2e^{-x} + e^{-x}(1-x)\left({\rm Ei}(x) - \log\frac{4x^2}{y}\right)\right]\,,
    \\
    \notag& I^{(5)}_a(x,y) = 32 x^3\left[3(1-2e^{-x})-x+4xe^{-x}+ (2-4x+x^2)e^{-x}\left({\rm Ei}(x) - \log\frac{4x^2}{y}\right)\right]\,,
    \\
    \notag& I^{(2)}_b(x,y) = 4x\left[-1+e^{-x}+xe^{-x}\left({\rm Ei}(x) - \log\frac{4x^2}{y}\right)\right]\,,
    \\
    & I^{(4)}_b(x,y) = 16x^2\left[-1+e^{-x}+x-3xe^{-x}-(-2+x)xe^{-x}\left({\rm Ei}(x) - \log\frac{4x^2}{y}\right)\right]\,.
\end{align}
With these notations, the antenna spectrum in the presence of a finite static medium, within the IOE approach, can be expressed in terms of the following contributions. Starting with the \textbf{Leading Order} terms, we have
\begin{align}
    \notag  \cJ^{\rm LO}_{\rm  in-in} &= 2\omega^2\Re\int_0^L dt_2 \int_0^{t_2}dt_1(1-\Delta_{\rm med}(t_1))e^{i\frac{\omega}{2}\delta\n^2 t_1(1+\frac{t_1}{T_{21}})}
    \\
    \notag &\hspace{0.5cm} \times \frac{e^{-\q_{21}^2/\hat P_{21}^2}}{\hat P_{21}^2 S_{21}^3}\Bigg[-2S_{21} + i\omega\delta\n^2 t_1(-C_{21}t_1+S_{21})-\frac{4i\omega C_{21}}{\hat P_{21}^2}\left(1 - \frac{\q_{21}^2}{\hat P_{21}^2}\right)
    \\
   \label{eq:LOinin} & \hspace{1.5cm}+2\omega \frac{\delta\n\cdot\q_{21}}{\hat P_{21}^2}\Big(t_1-C_{21}(-C_{21}t_1+S_{21})\Big) \Bigg] + \rm sym. \, ,
    \\
     \label{eq:LOinout}\cJ^{\rm LO}_{\rm in-out}  &= -\frac{2\omega}{\bkappa^2}\Re\,i\int_0^L dt_1 \left(1-\Delta_{\rm med}(t_1)\right)e^{i\frac{\omega}{2}\delta\n^2 t_1(1+\frac{t_1}{T_{L1}})}\nn
	& \hspace{0.5cm}\times \frac{e^{-\q_{L1}^2/\hat P_{L1}^2}}{C_{L1}^2}\bkappa\cdot\Big(\q_{L1} + \frac{i\hat P_{L1}^2}{2}\delta\n(-C_{L1}t_1 + S_{L1})\Big)+ \rm{sym.} \, ,
    \\
    \label{eq:LOoutout}\cJ^{\rm LO}_{\rm out-out} &= 4\omega^2\frac{\bkappa\cdot\bar\bkappa} {\bkappa^2\bar\bkappa^2}\left(1-\Delta_{\rm med}(L)\right)\cos \left(\frac{\bkappa^2-\bar\bkappa^2}{2} L \right)\, .
\end{align}
While the three \textbf{Next to Leading Order} terms are given by
\begin{align}
	& \cJ_{\rm in-in}^{\rm NLO,\, b} = -\frac{\hat q_0\omega^2}{4}\Re\int_0^L dt_2 \int_0^{t_2}dt_1(1-\Delta_{\rm med}(t_1))e^{i\frac{\omega}{2}\delta\n^2 t_1(1+\frac{t_1}{T_{21}})}\frac{(L-t_2)}{\q_{21}^4S_{21}^3}\nn
	\notag& \hspace{0.5cm}\times\Bigg[\Big(-2S_{21} + i\omega\delta\n^2 t_1(-C_{21}t_1+S_{21})\Big)I_a^{(3)}\left(\frac{\q_{21}^2}{\hat P_{21}^2},\frac{\q_{21}^2}{Q_b^2}\right)-i\frac{\omega C_{21}}{\q_{21}^2}I_a^{(5)}\left(\frac{\q_{21}^2}{\hat P_{21}^2},\frac{\q_{21}^2}{Q_b^2}\right)\\
	&  \hspace{2cm}+\omega \frac{\q_{21}\cdot\delta\n}{\q_{21}^2}\Big(t_1-C_{21}(-C_{21}t_1+S_{21})\Big)I_b^{(4)}\left(\frac{\q_{21}^2}{\hat P_{21}^2},\frac{\q_{21}^2}{Q_b^2}\right) \Bigg] + \rm sym. \, ,
\end{align}
\begin{align}
	&  \cJ_{\rm in-in}^{\rm NLO,\, r}  = -\frac{\hat q_0\omega^4}{2}\Re\int_0^L dt_2 \int_0^{t_2}dt_1(1-\Delta_{\rm med}(t_1))\int_{t_1}^{t_2}ds\,e^{i\frac{\omega}{2}\delta\n^2 t_1(1+\frac{t_1}{T_{s1}})}\frac{e^{-\bkappa^2/\hat P_{2s}^2}}{\hat P_{2s}^4S_{2s}^2S_{s1}^2\q_{2s1}^4}\nn
	& \hspace{0.5cm}\times\Bigg[2iC_{2s}(-C_{s1}t_1 + S_{s1})\bkappa\cdot\delta\n\, I_a^{(3)}\left(\frac{\q_{2s1}^2}{G_{2s1}},\frac{\q_{2s1}^2}{Q_r^2}\right)+\frac{Q_s^2(L,t_2)}{\q_{2s1}^2}I_a^{(5)}\left(\frac{\q_{2s1}^2}{G_{2s1}},\frac{\q_{2s1}^2}{Q_r^2}\right) \nn
	& \hspace{1cm}-i (2iC_{2s}\bkappa + Q_s^2(L,t_2)(-C_{s1}t_1 + S_{s1})\delta\n)\cdot \frac{\q_{2s1}}{\q_{2s1}^2}I_b^{(4)}\left(\frac{\q_{2s1}^2}{G_{2s1}},\frac{\q_{2s1}^2}{Q_r^2}\right)\Bigg]+ \rm sym. \, ,
\end{align}
\begin{align}
	& \cJ_{\rm in-out}^{\rm NLO,\, r}  = \frac{\hat q_0\omega^3}{2\bkappa^2}\Re\int_0^{L}dt_1(1-\Delta_{\rm med}(t_1))\int_{t_1}^L ds\,e^{i\frac{\omega}{2}\delta\n^2 t_1(1+\frac{t_1}{T_{s1}})}\nn
	\notag & \hspace{1cm}\times \frac{e^{-i\frac{T_{Ls}}{2\omega}\bkappa^2}}{C_{Ls}S_{s1}^2\q_{Ls1}^4}\Bigg[-i\frac{\bkappa\cdot\q_{Ls1}}{\q_{Ls1}^2}I_b^{(4)}\left(\frac{\q_{Ls1}^2}{G_{Ls1}},\frac{\q_{Ls1}^2}{Q_r^2}\right)
    \\
    & \hspace{4cm}+ \bkappa\cdot\delta\n(-C_{s1}t_1 + S_{s1})I_a^{(3)}\left(\frac{\q_{Ls1}^2}{G_{Ls1}},\frac{\q_{Ls1}^2}{Q_r^2}\right)\Bigg]+ \rm sym. \, .
\end{align}
\section{Numerical results}~\label{sec:numerics}
\begin{figure}[H]
     \centering
     \begin{subfigure}[h]{\textwidth}
         \centering
         \includegraphics[width=0.7\textwidth]{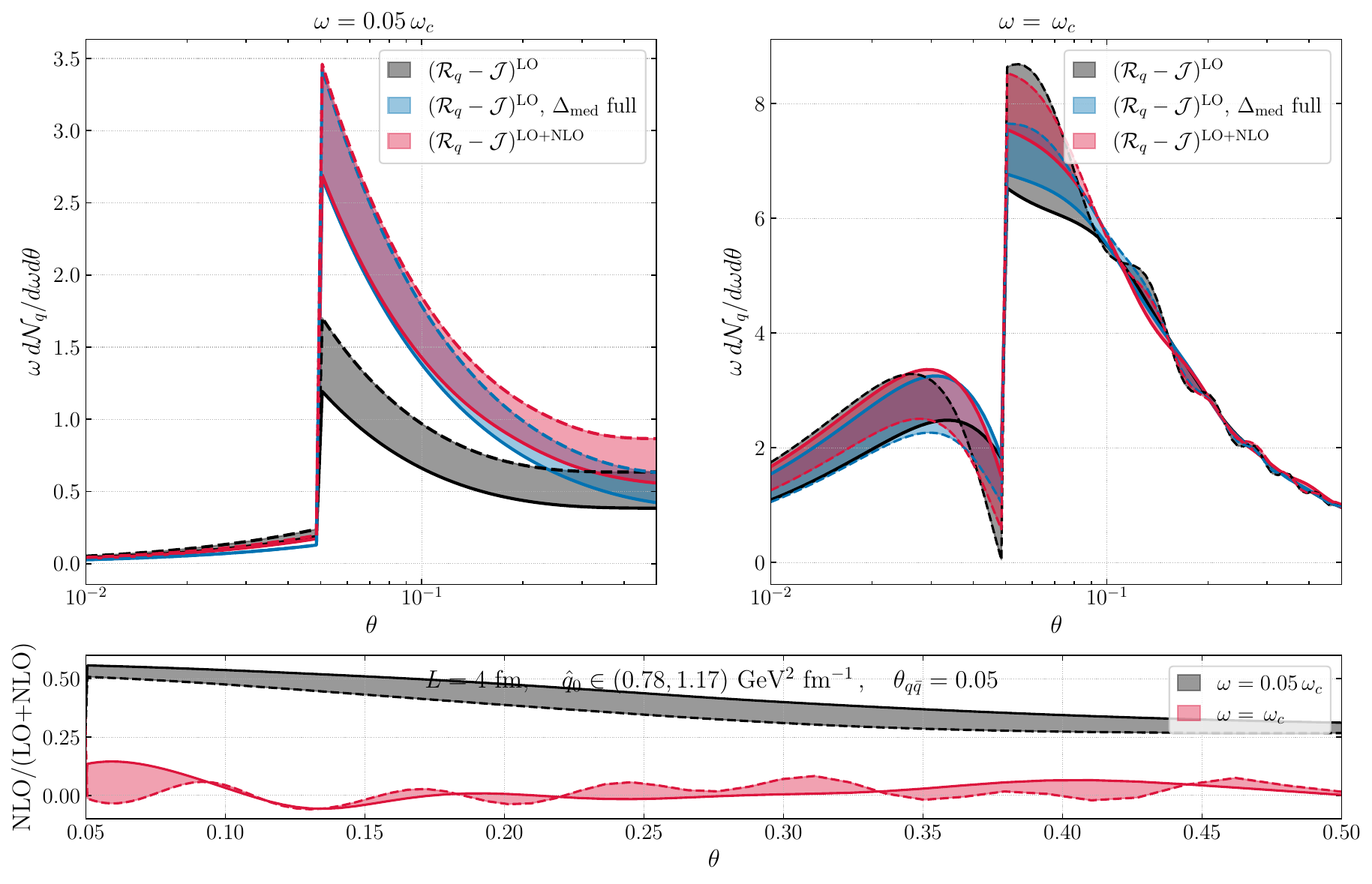}
     \end{subfigure}
     \begin{subfigure}[h]{\textwidth}
         \centering
         \includegraphics[width=0.7\textwidth]{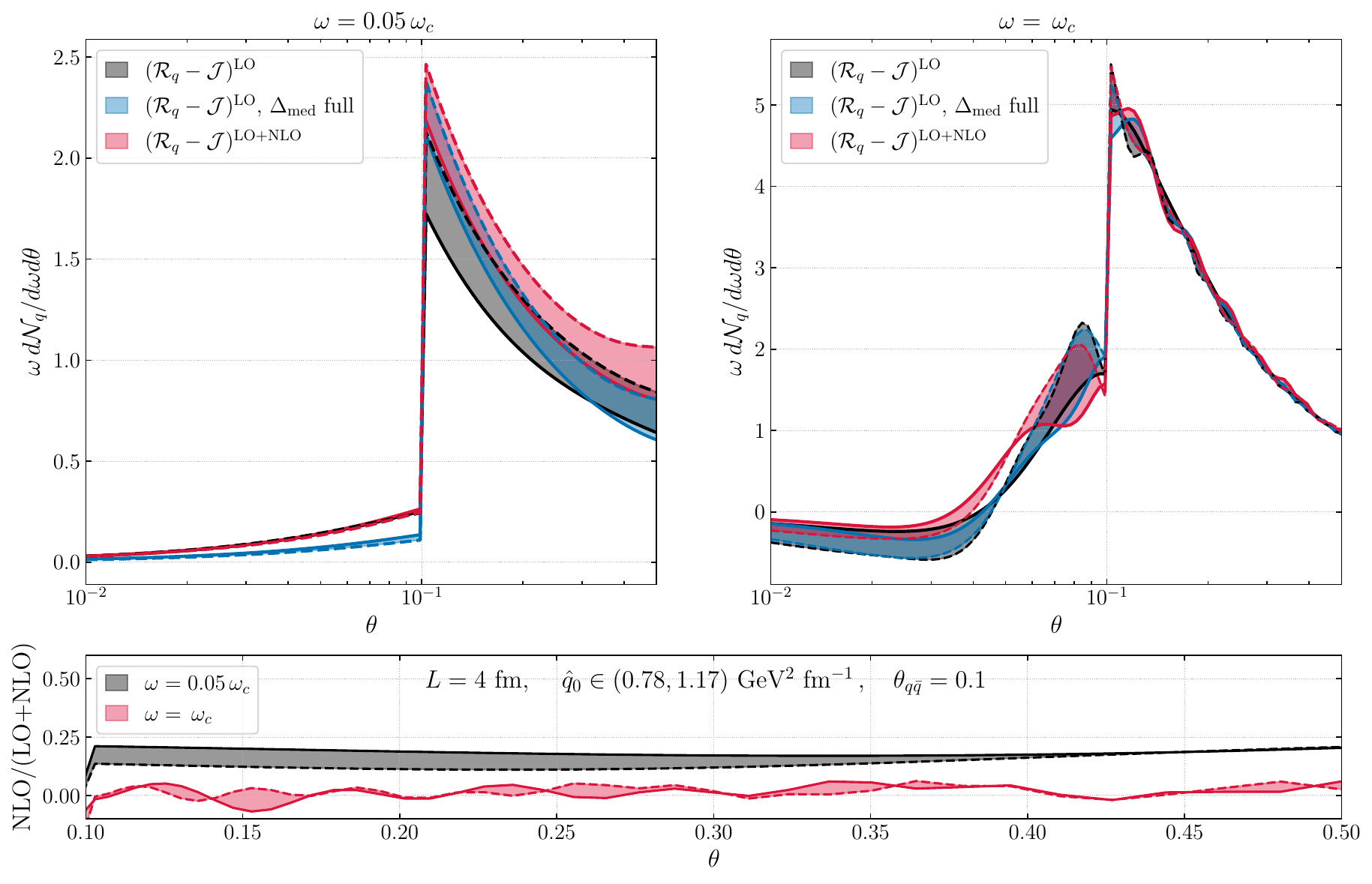}
     \end{subfigure}
     \begin{subfigure}[h]{\textwidth}
         \centering
         \includegraphics[width=0.7\textwidth]{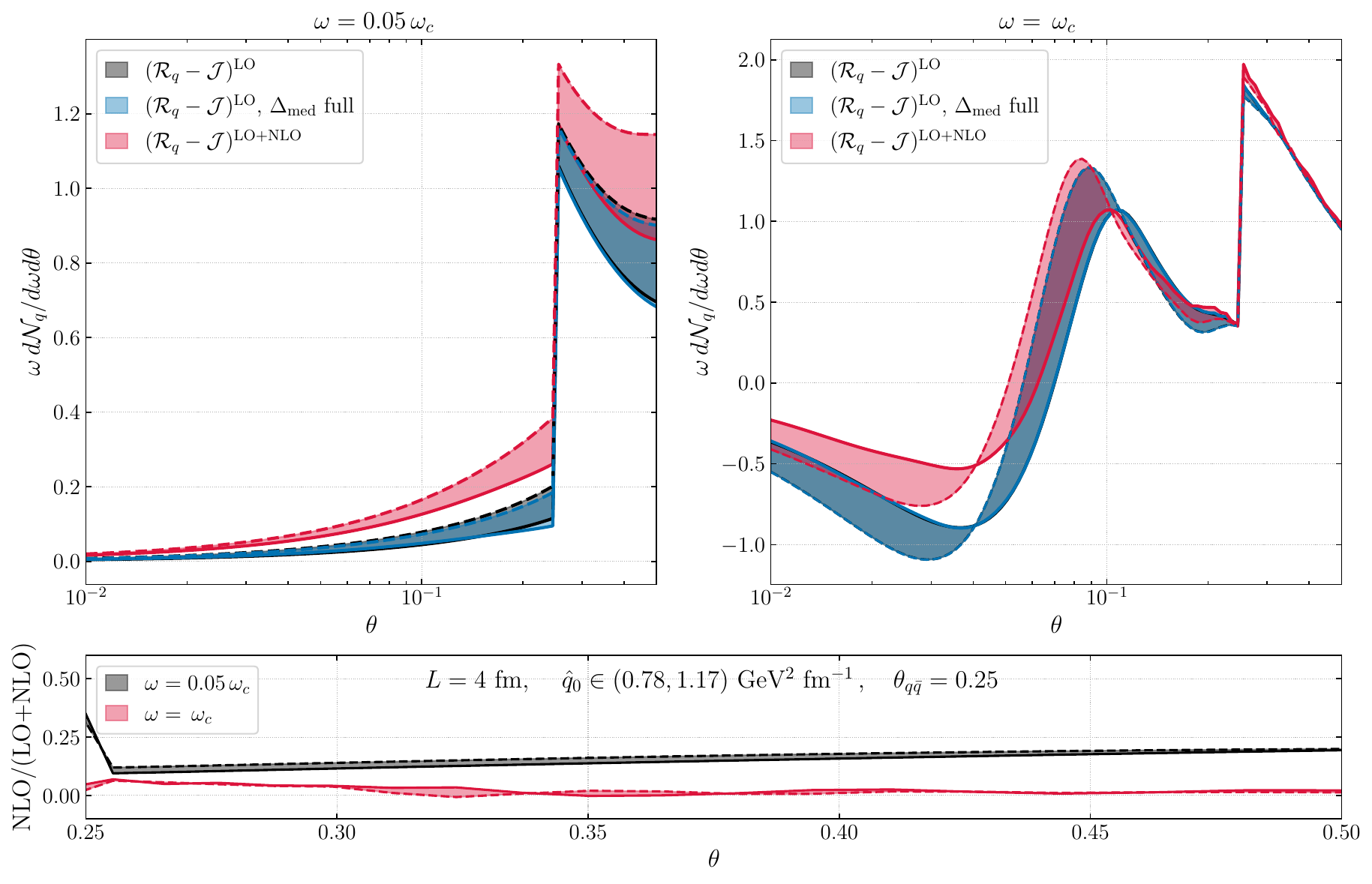}
     \end{subfigure}
     \caption{Antenna radiation spectrum at LO with $\Delta_{\rm med}$ calculated in the harmonic approximation, choosing the matching scale such that $\log{Q_{\Delta}^2/\mu_{\ast}^2}=1$ (\textbf{black}), at LO with $\Delta_{\rm med}$ calculated using the full dipole potential as in Eq.~\eqref{eq:DipolePotentialGW} (\textbf{blue}) and at LO+NLO (\textbf{red}). Each pair of panels corresponds to a different value of $\theta_{q\bar q}$, the left one to $\omega=0.05,\omega_c$ and the right one to $\omega = \omega_c$. The plot below each pair of panels shows the ratio between the NLO correction and the LO+NLO result.}
     \label{fig:3bands_NLO_w1wc}
\end{figure}

\indent In this section, we present the results from the numerical evaluation of the analytical expressions presented in Section~\ref{sec:FinalResults}. For this purpose, we take $L = 4$ fm, $\mu_{\ast} = 0.355$ GeV and $\hat q_0 \in (0.78,1.17)$ GeV$^2$ fm$^{-1}$, where we use a band to capture the dependence on $\hat q_0$, see Fig.~\ref{fig:3bands_NLO_w1wc}. This choice gives a range of values for the  bare saturation scale, varying between $Q_{s0} \in (1.75, 2.16)$ GeV. The gluon energy is chosen to be either $\omega = 0.05\,\omega_c$ or $\omega = \omega_c$, such that we can gauge the small and large energy behavior. We also explore a number of different antenna opening angles, which run through the values $\theta_{q\bar q}\in (0.05,0.1,0.25)$, corresponding to $r_{\perp}^{-1} \in (1,0.5,0.25)$. Notice that this set of parameters lies in the decoherence regime of the problem ($r_{\perp}^{-1} < Q_{s0}$), as defined in~\cite{Mehtar-Tani:2012mfa}. Following this earlier work, we chose to focus on this regime as it provides a transition between color-coherent and color-incoherent spectra due to the interplay between direct and interference terms. The complementary dipole regime is dominated by the interference contribution to the spectrum, showing only partial decoherence and thus having a suppressed direct contribution, see~\cite{Mehtar-Tani:2012mfa} for more details.

\begin{figure}[h!]
     \centering
     \begin{subfigure}[h]{0.45\textwidth}
         \centering
         \includegraphics[width=0.9\textwidth]{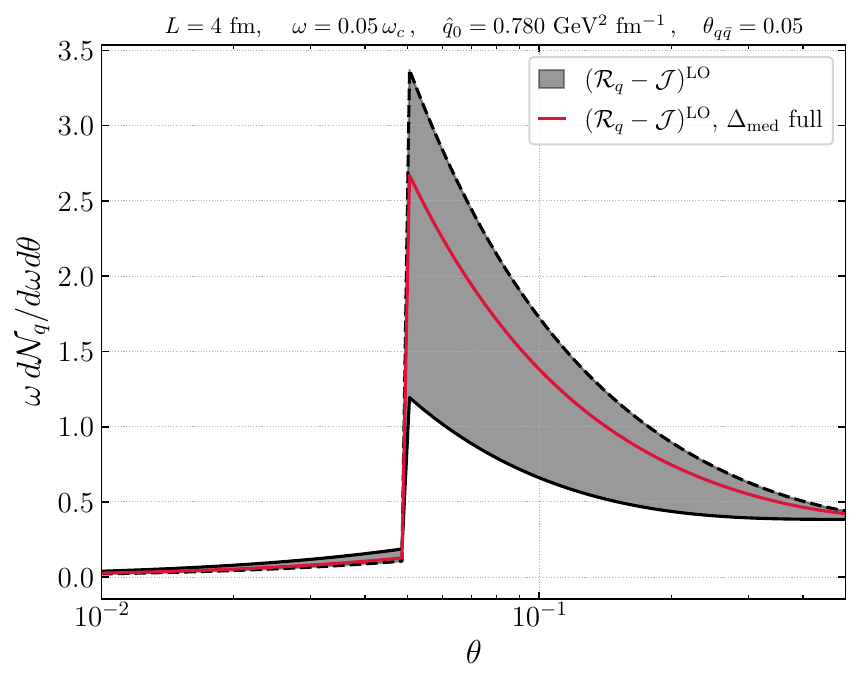}
     \end{subfigure}
     \begin{subfigure}[h]{0.45\textwidth}
         \centering
         \includegraphics[width=0.9\textwidth]{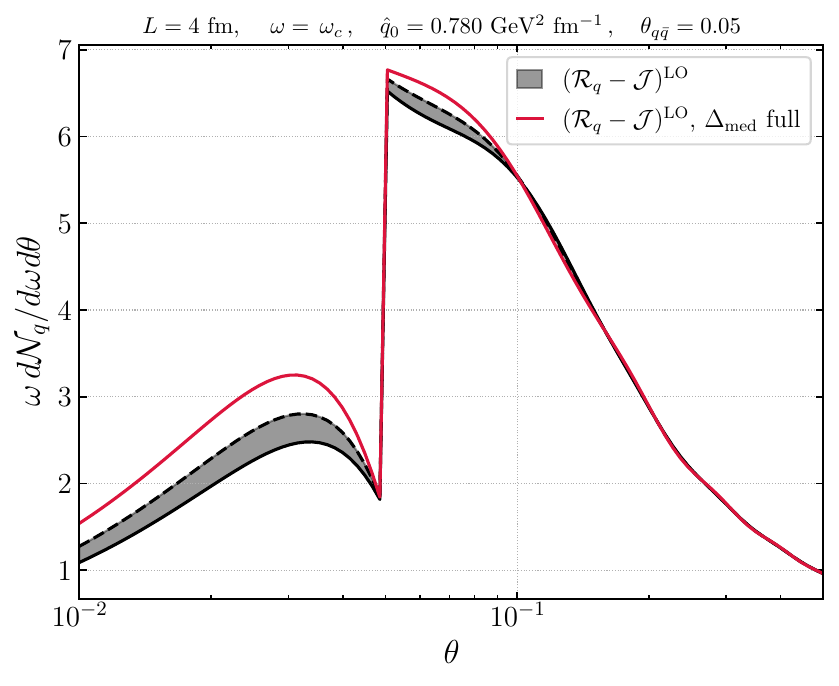}
     \end{subfigure}
     \begin{subfigure}[h]{0.45\textwidth}
         \centering
         \includegraphics[width=0.9\textwidth]{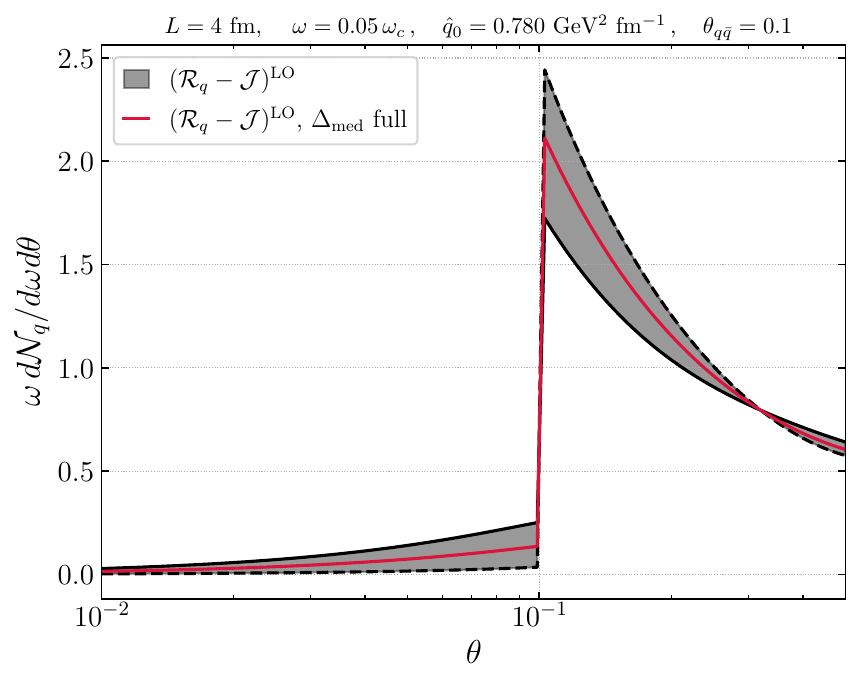}
     \end{subfigure}
     \begin{subfigure}[h]{0.45\textwidth}
         \centering
         \includegraphics[width=0.9\textwidth]{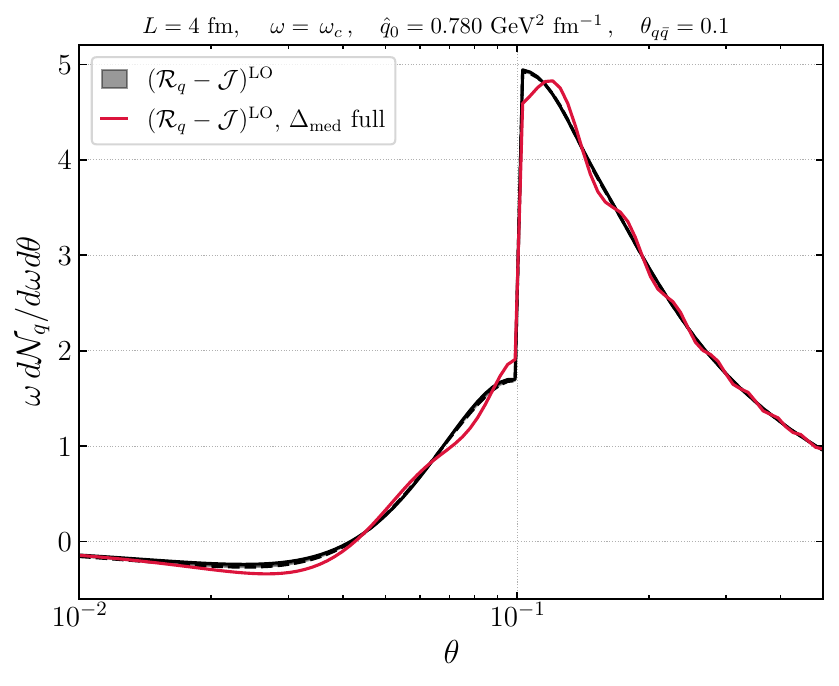}
     \end{subfigure}
     \begin{subfigure}[h]{0.45\textwidth}
         \centering
         \includegraphics[width=0.9\textwidth]{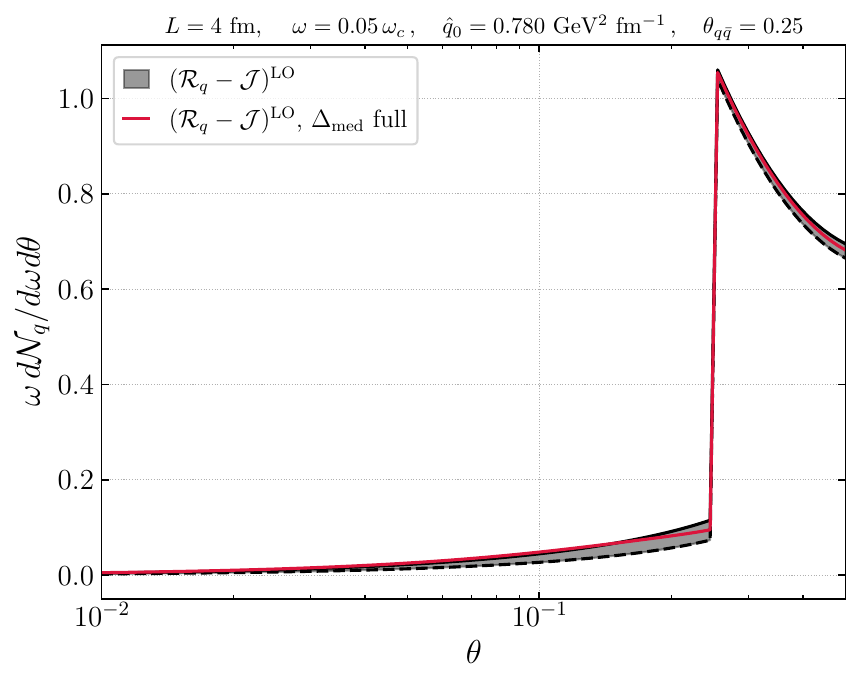}
     \end{subfigure}
     \begin{subfigure}[h]{0.45\textwidth}
         \centering
         \includegraphics[width=0.9\textwidth]{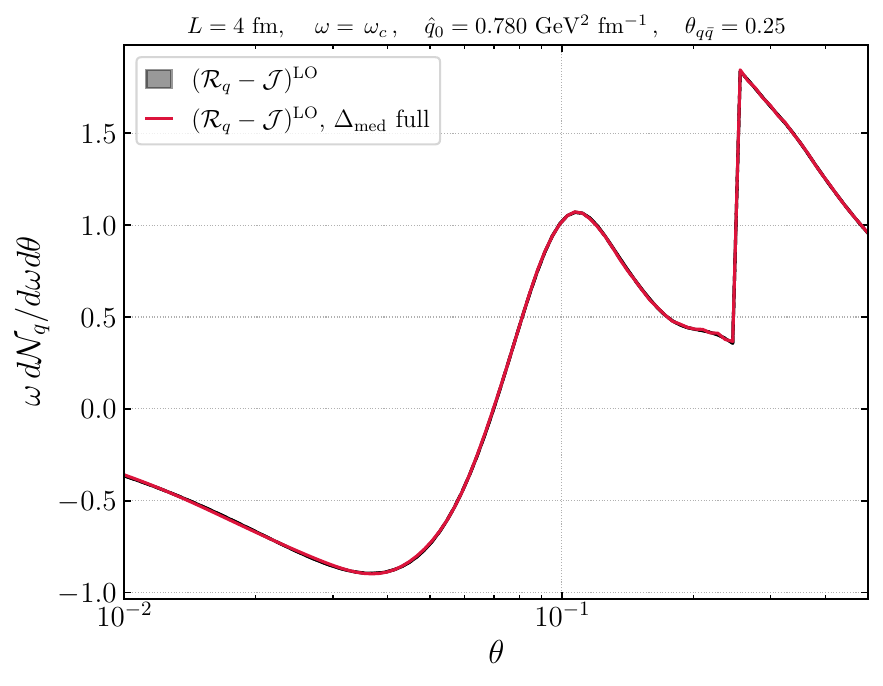}
     \end{subfigure}
     \caption{Antenna radiation spectrum at LO, with $\Delta_{\rm med}$ calculated using the harmonic oscillator approximation (black curves) and using the full dipole potential in Eq.~\eqref{eq:DipolePotentialGW} (red curve). The two black curves correspond to different choices of matching scale in the expansion of the decoherence factor $\Delta_{\rm med}$ to leading-order (Eq.~\eqref{eq:decoherence_factor_expansion}) ---  the solid line has $\log(Q_1^2/\mu_{\ast}^2) = 1$ and the dashed line has $Q_2 = 5\,Q_1$. A shaded band is drawn between the solid and dashed curves. Each pair of panels corresponds to a different value of $\theta_{q\bar q}$, the left one to $\omega=0.05,\omega_c$ and the right one to $\omega = \omega_c$.}
     \label{fig:LO_comp}
\end{figure}

As mentioned in Section~\ref{sec:AntennaIOE}, at NLO in the IOE, we can always retain the full form of $\Delta_{\rm med}$ using the potential in Eq.~\eqref{eq:DipolePotentialGW}, since its closed form can be evaluated analytically. We compared the full form of the coherence factor with the harmonic approximation expression in Fig.~\ref{Fig:DecoherenceFactor}. In terms of the IOE, keeping the decoherence factor in this form introduces corrections beyond NLO order to the final result, which are in principle negligible in comparison to the LO+NLO result. Keeping the full form of the potential avoids uncertainties that might arise from expanding $(1 - \Delta_{\rm med})$, such as the introduction of a new matching scale, $Q_{\Delta}$\footnote{We briefly explore an approach to define $Q_{\Delta}$ and discuss the possible physical scales involved in this choice in Appendix~\ref{app:Q_delta}.}. 
However, to have a LO result for the spectrum which one can compare to, it is necessary to make a choice for $Q_{\Delta}$, which corresponds to the LO result presented in Figs.~\ref{fig:LO_comp} and \ref{fig:3bands_NLO_w1wc}. 

In Fig.~\ref{fig:LO_comp} we start by comparing the LO spectrum obtained by calculating the decoherence factor $\Delta_{\rm med}$ with the full potential as defined in Eq.~\eqref{eq:DipolePotentialGW} and with the harmonic oscillator potential defined with a logarithmic dependence on a matching scale $Q_{\Delta}$, i.e.
\begin{align}\label{eq:decoherence_factor_expansion}
    \cV(\z) = \frac{1}{4}\hat q_0 \log\left(\frac{Q_{\Delta}^2}{\mu_{\ast}^2}\right)\z^2 \, .
\end{align}
 The solid black lines correspond to $\log\left(\frac{Q_{\Delta,1}^2}{\mu_{\ast}^2}\right) = 1$ while the dashed ones correspond to $Q_{\Delta,2} = 5\,Q_{\Delta ,1}$. The shaded band then corresponds to a variation of the matching scale $Q_{\Delta}$ between those two values. Firstly, we observe that the higher the energy, the weaker the dependence on the matching scale $Q_{\Delta}$, and the more insensitive the spectrum becomes to different choices of scattering potential in the calculation of $\Delta_{\rm med}$. This statement also holds the larger the value of $\theta_{q\bar q}$, an observation which is in agreement with the naive analysis of the decoherence factor shown in Fig.~\ref{Fig:DecoherenceFactor}. Nevertheless, for $\theta_{q\bar q} \lesssim 0.1$ and for moderate values of $\omega$, the spectrum is quite sensitive to the choice of the matching scale $Q_{\Delta}$. In this region of parameter space, we conclude that, in the absence of a well-defined matching condition for $Q_{\Delta}$, using the harmonic oscillator approximation for the decoherence factor introduces substantial uncertainty into the antenna spectrum.

Retaining the full form of the potential, as given in Eq.~\eqref{eq:DipolePotentialGW}, does not require defining $Q_{\Delta}$, thus avoiding the uncertainty associated with this matching scale. However, in this case, it requires including the NLO corrections to the broadening and radiative kernels in order to maintain full consistency and controlled accuracy in the final spectrum.  To this end, in Fig.~\ref{fig:3bands_NLO_w1wc}, we present the antenna radiation spectrum at LO and LO+NLO, with bands showing a variation of the bare quenching parameter $\hat q_0 \in (0.78, 1.17)$. The left panels have $\omega = 0.05\,\omega_c$ while the right ones have $\omega = \,\omega_c$ and the bottom plot presents the ratio of the NLO correction to the full NLO+LO result. For the LO spectrum in the black band the decoherence factor $\Delta_{\rm med}$ is calculated in the harmonic oscillator approximation with a specific choice of matching scale $Q_{\Delta}$. For the spectrum in the blue band, the only correction comes from retaining the full dipole potential in the decoherence factor $\Delta_{\rm med}$. The potentials entering the emission and broadening kernels are kept in the HO approximation. The LO+NLO result includes all NLO corrections, including the full potential in $\Delta_{\rm med}$ as well as the kernel corrections calculated in the IOE, which we summarized in Section~\ref{sec:FinalResults}. The comparison in Fig.~\ref{fig:LO_comp} demonstrated a strong dependence of the LO spectrum on the choice of the matching scale $Q_{\Delta}$, especially for lower energies and smaller opening angles. Nevertheless, it is still relevant to draw a comparison to the LO result (black band) which is more frequently used in the literature, i.e., fixing the matching scale as $\log\left(\frac{Q_{\Delta}^2}{\mu_{\ast}^2}\right)=1$ (solid lines in Fig.~\ref{fig:LO_comp}). The blue band, on the other hand, avoids this uncertainty, providing a more precise estimation of the NLO corrections to the broadening and radiation kernels when compared to the red band.

We begin by assessing the magnitude of the full set of NLO corrections, i.e., we focus on comparing the black band (LO) with the red band (LO+NLO). First, for $\theta > \theta_{q\bar q}$, looking at the three ratio plots (corresponding to the three different opening angles $\theta_{q\bar q}\in (0.05,0.1,0.25)$), one sees that the corrections get larger the smaller the $\theta_{q\bar q}$. Additionally, there is no overall trend with $\theta$, since the corrections grow with $\theta$ for $\theta_{q\bar q} = 0.05$ but have the opposite behavior for $\theta_{q\bar q} = 0.25$. In this region of $\theta > \theta_{q\bar q}$, the NLO terms are especially pronounced for $\theta \sim \theta_{q\bar q}$ and for $\omega = 0.05\,\omega_c$, where they can reach $\sim 50\%$. As one increases the energy to $\omega = \omega_c$, the corrections become less pronounced and they decrease as the emission angle increases. This is in apparent contradiction what one would expect from large $\k/Q_s$ and large $\omega/\omega_c$ emissions being dominated by the effect of a single hard scattering. However, such a clear behaviour is to be expected in the assymptotic limit of $\omega \gg \omega_c$ and $\k \gg Q_s$ --- in the GLV limit --- which is not fully captured here. For the remaining of the angular region ($\theta < \theta_{q\bar q}$), the corrections seem substantial for the largest energy $\omega = \omega_c$ or for the largest opening angle $\theta_{q\bar q} = 0.25$. In particular, for $\omega=\omega_c$ and $\theta_{q\bar q} = 0.05$, the dependence on $\hat q_0$ seems to be exactly reversed when accounting for the NLO corrections.

We now turn to the analysis of the magnitude of the NLO corrections to the radiative and broadening kernels, i.e.,  we compare the blue band (LO with $\Delta_{\rm med}$ calculated with the full dipole potential in Eq.~\eqref{eq:DipolePotentialGW}) with the red band (LO+NLO). First, for $\theta > \theta_{q\bar q}$, we see that for the smallest energy (left panels), the magnitude of NLO corrections to the kernels increases with increasing angle $\theta$, consistently accounting for an enhancement of the spectrum in the region $\theta > \theta_{q\bar q}$, which can go up to about $30\%$. In fact, it is clear that as one dials $\theta$ or $\theta_{q\bar q}$ up, the NLO corrections get increasingly dominated by the corrections to the kernels rather than to $\Delta_{\rm med}$. This is clear for $\theta_{q\bar q} = 0.25$, where the blue and black bands are exactly overlapped. Otherwise, for $\theta \gtrsim \theta_{q\bar q}$ and for small enough $\theta_{q\bar q}$, modifying $\Delta_{\rm med}$ is responsible for the largest correction. As we increase the energy to $\omega_c$, we see that for $\theta \gtrsim \theta_{q\bar q}$ and $\theta_{q\bar q} < 0.25$ there is an increase in the magnitude of NLO corrections to the kernels, which are of the order of $10\%$. In this region, modifying $\Delta_{\rm med}$ is not enough to qualitatively describe the enhancement of radiation. For $\theta < \theta_{q\bar q}$ and for $\omega = \omega_c$, the corrections to $\Delta_{\rm med}$ seem to be enough to predict the correct $\hat q_0$ dependence for $\theta_{q\bar q} = 0.05$. For the remaining opening angle values the blue and black bands overlap, demonstrating insensitivity to $\Delta_{\rm med}$. In fact, in this region it is the direct term that dominates, so the NLO corrections come mostly from the kernels entering its definition.

\begin{figure}[H]
     \centering
     \begin{subfigure}[h]{0.48\textwidth}
         \centering
         \includegraphics[width=0.9\textwidth]{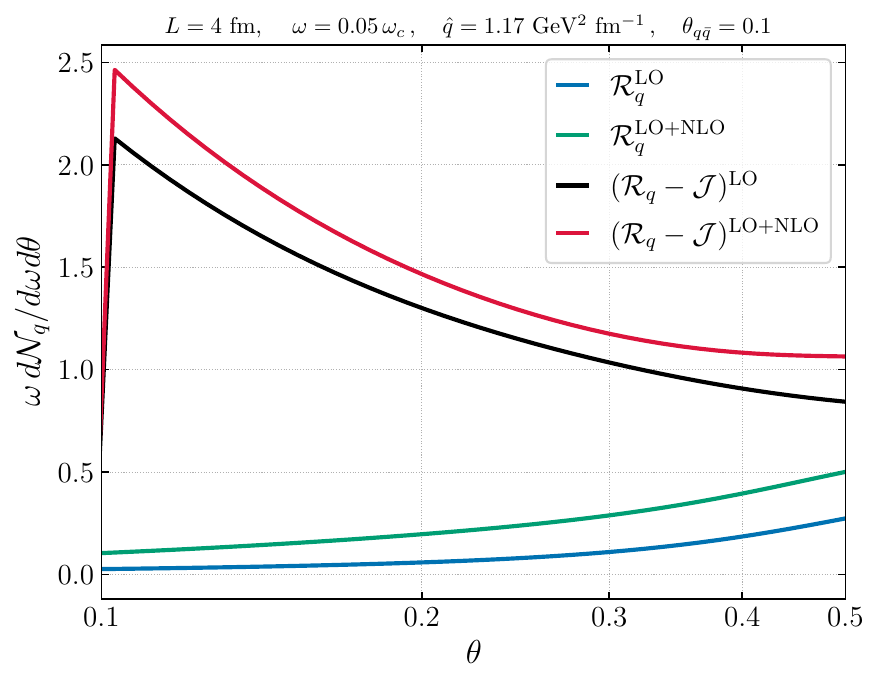}
     \end{subfigure}
     \begin{subfigure}[h]{0.48\textwidth}
         \centering
         \includegraphics[width=0.9\textwidth]{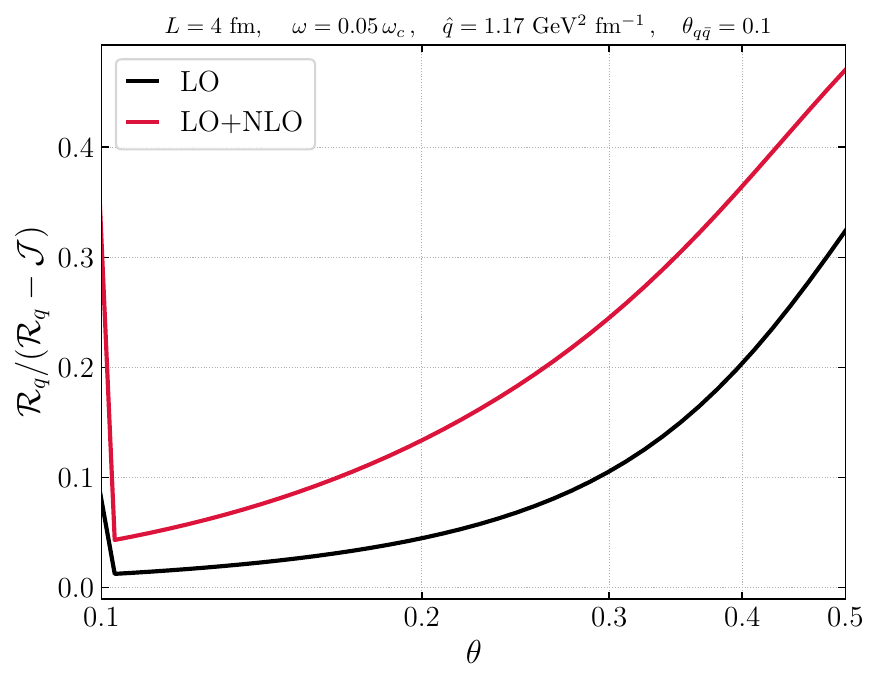}
     \end{subfigure}
     \caption{\textit{Left:} Antenna radiation spectrum at LO with $\Delta_{\rm med}$ calculated in the harmonic approximation (\textbf{black}) and at LO+NLO (\textbf{red}) for $\theta > \theta_{q\bar q}$, for fixed $\theta_{q\bar q} = 0.1$. The direct terms at LO (\textbf{blue}) and LO+NLO (\textbf{green}) are also plotted for direct comparison. \textit{Right:} Ratio between the direct term and the full spectrum at LO and LO+NLO.}
     \label{fig:decoherence_plot}
\end{figure}

Finally, it is interesting to study whether the color decoherence of the $q\bar q$ pair due to interactions with the medium is slowed down or sped up when including corrections beyond the harmonic oscillator potential. To do so, one can examine how similar the full spectrum is to the independent/direct term, which, in specific gauges, can be interpreted as representing the incoherent emission of the gluon off the quark. In the left panel of Fig.~\ref{fig:decoherence_plot}, we plot the LO and LO+NLO spectra similar to what we presented in Fig.~\ref{fig:3bands_NLO_w1wc} but for a single value of $\theta_{q\bar q} = 0.1$, $\hat q_0 = 0.78$ GeV$^2$ fm$^{-1}$ and for emission angles in the relevant region $\theta > \theta_{q\bar q}$. For direct comparison, we also plot the direct contribution to the spectrum for both orders. Clearly, both spectra tend to their independent counterparts the larger the emission angle $\theta$. When looking at the right panel, where we plot the ratio between the direct term and the full spectrum, we see that when including NLO corrections the ratio is not only larger but it is also increasing faster than the LO ratio for most of the angular range. This suggests that the loss of color coherence between the quark and anti-quark is sped up when going beyond the harmonic oscillator approximation.

\section{Conclusion and Outlook}~\label{sec:conclusion}
\indent In this work we have  extended the Improved Opacity Expansion approach, which was previously studied for broadening~\cite{Barata:2020rdn} and radiation off a single quark~\cite{Mehtar-Tani:2019tvy,Mehtar-Tani:2019ygg,Barata:2020sav,Barata:2021wuf}, for the case of in-medium gluon emission off a color-singlet $q\bar q$ antenna. For a finite static medium, we calculated the next-to-leading order corrections to the interference term defined in Eq.\eqref{eq:InterferenceDef}. Taking the vanishing dipole size limit, $\delta\n \rightarrow 0$, we then extracted the corresponding NLO corrections to the direct term in Eq.\eqref{RqDef}. Comparing these results with those obtained for gluon emission off a single quark in~\cite{Barata:2021wuf} provided an important consistency check of our analytical calculations. 
The obtained spectrum qualitatively gauges the features of both single hard scattering and multiple soft scattering regimes, expanding the existing works on the in-medium QCD antenna radiation pattern.

In Section~\ref{sec:FinalResults} we summarized the analytical results for the double differential spectrum and in Section~\ref{sec:numerics} we provided a numerical analysis which focuses on the decoherence regime presented in~\cite{Mehtar-Tani:2012mfa}. In this section, we studied the corrections to the spectrum associated with accounting for the full dipole potential in~\eqref{eq:DipolePotentialGW}, compared to the HO approximation (see Fig.~\ref{fig:LO_comp}). We found that, for lower energies and smaller opening angles, a modest variation of the matching scale $Q_{\Delta}$ brings a large uncertainty into the spectrum. This uncertainty can be avoided by directly using the full dipole potential and, for consistency, the remaining corrections to the emission and broadening kernels need also to be included. In Fig.~\ref{fig:3bands_NLO_w1wc}, we compared the final spectrum calculated at the LO solely using the HO approximation with the LO+NLO result for a range of $\hat{q}_0$.
We found that the corrections are substantial across the full range of discussed parameters, peaking at $\sim 50\%$ of the full spectrum for the smallest energy and opening angle. We also discovered that, while the contributions associated with the decoherence factor $\Delta_{\rm med}$ dominate the NLO corrections at smaller $\theta_{q\bar q}$ and especially for emissions angles $\theta \sim \theta_{q\bar q}$, the broadening and emission kernels dominate the NLO corrections at larger $\theta_{q\bar q}$. Finally, in Fig.~\ref{fig:decoherence_plot}, we plotted the ratio between the direct term of the spectrum, which can be interpreted as representing the incoherent radiation off the quark, and the full antenna spectrum. Including NLO corrections to the spectrum results in substantially closer to $1$, suggesting that the onset of decoherence of the $q\bar q$ is accelerated when accounting for the effect of a single hard scattering.

A natural consistency check would involve considering the GLV limit of the full spectrum to analytically identify the SH regime and compare it with results from, e.g.,~\cite{Mehtar-Tani:2011vlz}. However, such an exercise is rather challenging due to the interplay of multiple scales and the complicated analytical form of the NLO corrections. Numerically, it is also challenging to fully quantify the impact of the single-hard regime, since that would require reaching the asymptotic limit of GLV. We leave such endeavors for future work. 

The results presented here contribute to the ongoing development of the theory of jet–medium interactions and offer several possibilities for further extension. Our results can be generalized to the case of an evolving medium profile, following the approaches of, e.g.,~\cite{Salgado:2003gb,Arnold:2008iy,Adhya:2019qse,Caucal:2020uic}, which would lead to a more intricate structure for the $S_{12}$ and $C_{12}$ functions. It would also be interesting to explore the interplay of the obtained corrections with recent related advancements in the description of radiative processes in the medium, see e.g.~\cite{Abreu:2024wka, Pablos:2024muu}, or gauge the impact of these corrections in calculations of energy correlators in HICs (see e.g.~\cite{Andres:2022ovj,Barata:2023zqg,Yang:2023dwc,Barata:2023bhh,Barata:2024bmx,Barata:2024wsu,Barata:2024ieg,Apolinario:2025vtx,Barata:2025fzd,Moult:2025nhu}). Furthermore, the results of this paper could be used as a theoretical ingredient of a Monte Carlo generator for partonic cascades in the medium.

\section*{Acknowledgments} 

The authors are grateful to Liliana Apolinário, João Barata, André Cordeiro, Xoán Mayo Lopez, Guilherme Milhano, Andrey Sadofyev, Carlos Salgado and Alba Soto-Ontoso for useful discussions and comments on the manuscript. In particular, the authors would like to thank João Barata for multiple insightful discussions throughout the development of this project. The work of MVK is partially funded by the grant $\#$ 24-2-1-69-1 of the Foundation for the Advancement of Theoretical Physics “BASIS”. The work of JMS has been supported by MCIN/AEI (10.13039/501100011033) and ERDF (grant PID2022-139466NB-C21) and by Consejería de Universidad, Investigación e Innovación, Gobierno de España and Unión Europea – NextGenerationEU under grant AST22\_6.5.

\newpage
\appendix
\section    {The matching scale for the decoherence factor}\label{app:Q_delta}

In the main text we highlighted that retaining the full form of the decoherence factor in Eq.~\eqref{eq:CoherenceFDef} with the full potential in Eq.~\eqref{eq:DipolePotentialGW} is consistent with the IOE approach, in that it accounts for the LO+NLO result and introduces higher order corrections which one can in principle assume to be small.
Additionally, 
this avoids uncertainties associated with the ambiguity in the definition of the new "decoherence" matching scale $Q_{\Delta}$. 
However, we recognize the need to compare to a full LO result, e.g., the one derived under the multiple soft scattering approximation, to make contact with the seminal works on the study of in-medium color coherence of an antenna. In the small distance limit, the decoherence factor reads
%
\begin{align}
    1-\Delta_{\rm med}(t) & = \exp\Big[-\frac{1}{4}\hat q_0 \theta_{q\bar q}^2\int_0^t ds\,  s^2 \left( \log\frac{Q_{\Delta}^2}{\mu_{\ast}^2} + \log\frac{1}{Q_{\Delta}^2\theta_{q\bar q}^2\,s^2}\right)\Big]\,,
\end{align}
which at LO, i.e, ignoring the second term in the integrand, reads
\begin{align}\label{eq:delta_med_LO}
(1-\Delta_{\rm med} (t))^{\rm LO} = \exp\Big[-\frac{1}{12}\left(t/t_d\right)^3\Big]\,,
\end{align}
where we defined the decoherence time $t_d \equiv (\theta_{q \bar q}^2\,\hat{q}_0\log Q_{\Delta}^2/\mu_{*}^2)^{-1/3}$ (see e.g.~\cite{Mehtar-Tani:2012mfa} for a detailed discussion of this scale). Unlike the cases of the $Q_r$ and $Q_b$ matching scales in Eqs.~\eqref{eq:QrCondition} and~\eqref{eq:QbCondition}, there is no well-established $Q_{\Delta}$ currently known in the literature, so we discuss its defining condition below.

First, we recall that the qualitative shape of the antenna spectrum is strongly determined by the interplay of two characteristic medium-induced transverse scales (see e.g.~\cite{Mehtar-Tani:2012mfa}), particularly the inverse maximal dipole size $r_{\perp}^{-1} = (\theta_{q\bar q}L)^{-1}$ and the saturation scale $Q_{s0} = \sqrt{\hat q_0 L}$\,\footnote{In this consideration, we assume the vacuum scale $\delta \bkappa$ to be much smaller than the medium scales.}. 

The interplay between these two scales gives rise to two regimes, which can be expressed in terms of the bare decoherence time $t_{d0} = (\theta_{q \bar q}^2\,\hat{q}_0)^{-1/3}$. These are the the dipole regime, where $t_{d0} > L$, and the decoherence regime, where $t_{d0} < L$; both regimes were discussed in Section~\ref{sec:numerics}. Given the different limiting behaviours of $\Delta_{\rm med}$ in these two regimes, it is natural to consider a different matching scale for each. In the decoherence regime, the relevant scale can be identified with the typical inverse dipole size after the decoherence time, such that
\begin{align}\label{eq:Qd}
    Q_{\Delta}(\theta_{q\bar q}) = (\theta_{q\bar q}t_d)^{-1} = \left(\frac{\theta_{q\bar q}}{\hat q_0\log\frac{Q_{\Delta}^2(\theta_{q\bar q})}{\mu_{\ast}^2}}\right)^{-1/3} \equiv Q_d(\theta_{q\bar q})
\end{align}
For the dipole regime, on the other hand, one has $t_{d0} > L$ and another scale should be used. A suitable candidate is the inverse dipole size after the pair traverses the whole extent of the medium $Q_{\Delta}(\theta_{q\bar q}) = (\theta_{q\bar q} L)^{-1} \equiv Q_a(\theta_{q\bar q})$. 

\begin{figure}[H]
     \centering
     \begin{subfigure}[h]{0.48\textwidth}
         \centering
         \includegraphics[width=0.9\textwidth]{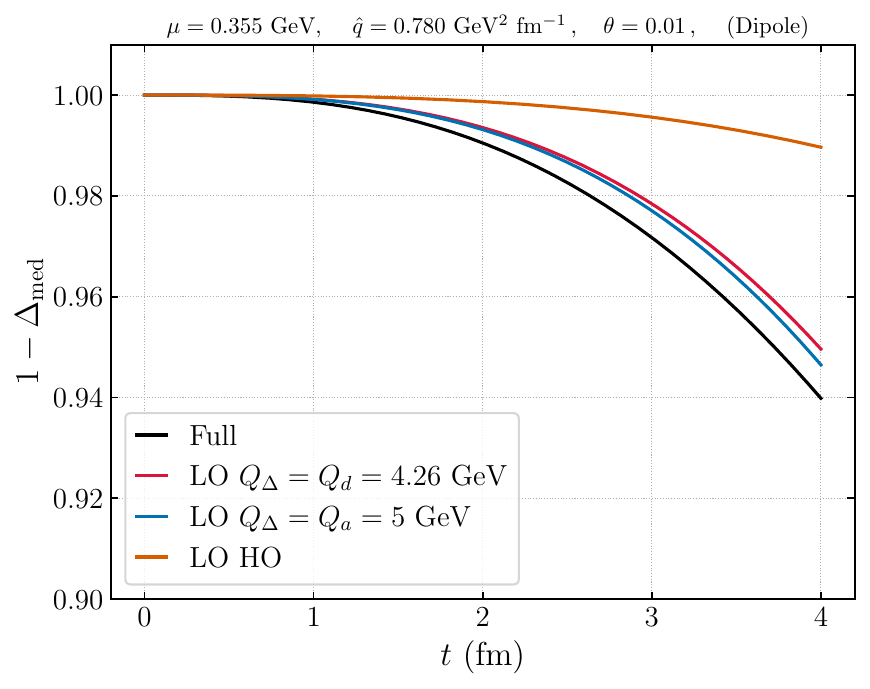}
     \end{subfigure}
     \begin{subfigure}[h]{0.48\textwidth}
         \centering
         \includegraphics[width=0.9\textwidth]{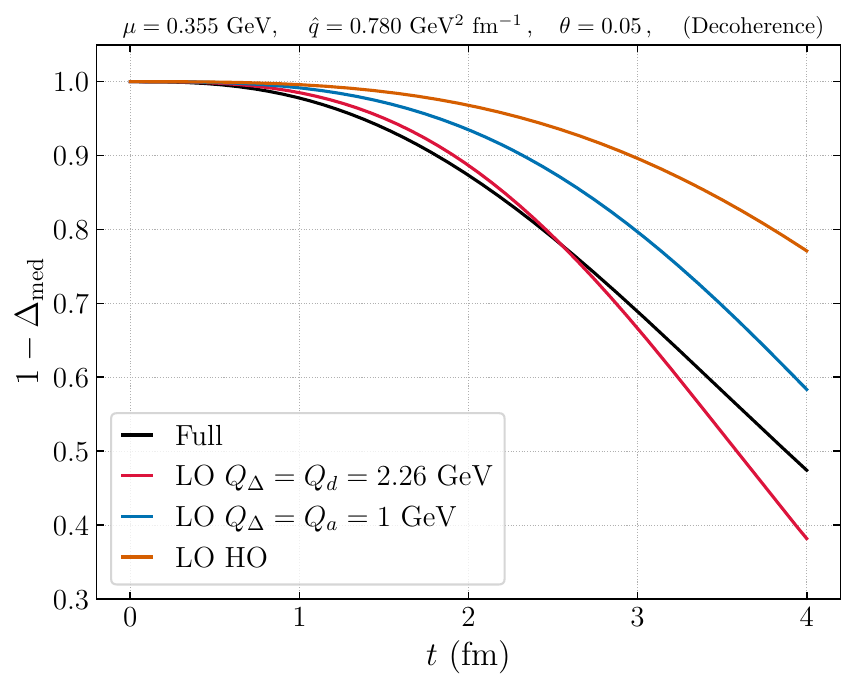}
     \end{subfigure}
     \caption{Decoherence factor $1-\Delta_{\rm med}$ calculated using the full potential in Eq.~\eqref{eq:DipolePotentialGW} (black), using the LO expression in Eq.~\eqref{eq:delta_med_LO} with $Q_{\Delta} = Q_d$ as defined in Eq.~\eqref{eq:Qd} (blue), with $Q_{\Delta} = Q_a = (\theta_{q\bar q}L)^{-1}$ (red) and with $\log\frac{Q_{\Delta}^2}{\mu_{\ast}^2}=1$ as is usual in the HO approximation (orange). The left panel corresponds to the dipole regime ($t_{d0} > L$) and the right panel corresponds to the decoherence regime ($t_{d0} < L$).}
     \label{fig:Q_delta}
\end{figure}

To explore whether this reasoning might be valid, in Fig.~\ref{fig:Q_delta} we show $(1-\Delta_{\rm med}(t))^{\rm LO}$ in Eq.~\eqref{eq:delta_med_LO} computed using the two different, physically motivated choices for the matching scale ($Q_{d}$ and $Q_a$). For completeness, we also show the decoherence factor calculated with the full potential as in Eq.~\eqref{eq:DipolePotentialGW} and with usual the harmonic approximation which naively sets $\log\frac{Q_{\Delta}^2}{\mu_{\ast}^2} = 1$. In the dipole regime ($t_{d0} > L$, left panel), the two choices of matching scale give similar results, with $Q_a$ being closest to the exact result. On the other hand, in the decoherence regime ($t_{d0} < L$, right panel), the two choices give sufficiently different magnitudes for the decoherence factor, with $Q_d$ resulting in a better agreement with the exact result. 

This example demonstrates that defining separate, regime-specific conditions for $Q_{\Delta}$ leads to better agreement between the leading-order (LO) approximation and the full potential result—significantly improving accuracy compared to the harmonic oscillator (HO) approximation. The defined scale directly affects the decoherence time $t_d$ and, in certain parametric regions, can modify it substantially. We expect this point to be relevant in studies that rely on the interplay of characteristic medium-induced scales, potentially offering deeper insight into the underlying physical processes.

\bibliographystyle{elsarticle-num}
\bibliography{references.bib}
\end{document}